# Monolayer black phosphorus by sequential wet-chemical surface oxidation

Stefan Wild,[a] Vicent Lloret,[a] Victor Vega-Mayoral,[b,c] Daniele Vella,[c,d] Edurne Nuin,[a,e] Martin Siebert,[f] Maria Koleśnik-Gray,[f] Mario Löffler,[g] Karl J. J. Mayrhofer,[g] Christoph Gadermaier,[c,h] Vojislav Krstić,[f] Frank Hauke,[a] Gonzalo Abellán*[a,e] and Andreas Hirsch*[a]

a. S. Wild, V. Lloret, Dr. E. Nuin, Dr. F. Hauke, Dr. G. Abellán, Prof. A. Hirsch
   Department of Chemistry and Pharmacy and Joint Institute of Advanced Materials and Processes (ZMP), Friedrich-Alexander-Universität Erlangen-Nürnberg (FAU), Nikolaus Fiebiger-Strasse 10, 91058 Erlangen and Dr.-Mack Strasse 81, 90762 Fürth, (Germany).
   E-mail: gonzalo.abellan@fau.de (G.A.) andreas.hirsch@fau.de (A.H.)
b. Dr. V. Vega-Mayoral
   CRANN & AMBER Research Centers, Trinity College Dublin, Dublin 2, Ireland & School of Physics, Trinity College Dublin, Dublin 2, Ireland
c. Dr. V. Vega-Mayoral, Dr Daniele Vella, Prof. C. Gadermaier
   Department for Complex Matter, Jozef Stefan Institute, Jamova 39, 1000 Ljubljana, Slovenia & Jozef Stefan International Postgraduate School, Jamova 39, 1000 Ljubljana, Slovenia
d. Dr. Daniele Vella
   Department of Physics National University of Singapore, 2 Science Drive 3, Singapore 117542, Singapore.
e. Dr. E. Nuin, Dr. G. Abellán
   Instituto de Ciencia Molecular (ICMol), Universidad de Valencia, Catedrático José Beltrán 2, 46980, Paterna, Valencia, Spain
f. M. Siebert, Dr. M. Koleśnik-Gray, Prof. V. Krstić
   Department of Physics, Friedrich-Alexander-Universität Erlangen-Nürnberg (FAU), Staudtstr. 7, 91058 Erlangen, Germany.
g. M. Löffler, Dr. K. J. J. Mayrhofer
   Helmholtz Institute Erlangen-Nürnberg for Renewable Energy (IEK-11), Forschungszentrum Jülich GmbH, Egerlandstraße 3, 91058 Erlangen, Germany, Department of Chemical and Biological Engineering, Friedrich-Alexander-Universität Erlangen-Nürnberg (FAU), Immerwahrstraße 2a, 91058 Erlangen, Germany
h. Prof. C. Gadermaier
   Department of Physics
   Politecnico di Milano, Piazza Leonardo da Vinci 32, 20133 Milano, Italy

**Abstract**

We report a straightforward chemical methodology for controlling the thickness of black phosphorus flakes down to the monolayer limit by layer-by-layer oxidation and thinning, using water as solubilizing agent. Moreover, the oxidation process can be stopped at will by two different passivation procedures, namely the non-covalent functionalization with perylene diimide chromophores, which prevents the photooxidation, or by using a protective ionic liquid layer. The obtained flakes preserve their electronic properties as demonstrated by fabricating a BP field-effect transistor (FET). This work paves the way for the preparation of BP devices with controlled thickness.

**Introduction**

Black Phosphorus (BP) is attracting tremendous attention as a new 2D material due to its excellent properties such as its direct bandgap (spans over 0.3–2 eV), high carrier mobility (up to 6000 $cm^2 \cdot V^{-1} \cdot s^{-1}$), good on/off ratio and exotic in-plane anisotropy, which make it unique for thermal imaging, thermoelectrics, sensing, fiber optics communication, and photovoltaics, to name only a few.[1–7] Pristine BP is very oxophilic and under ambient conditions easily converted to $PO_2^{3-}$, $PO_3^{3-}$, and $PO_4^{3-}$ species.[8–12] Water is crucial in the degradation process, indeed, water allows for the dissociation of oxidized P from the BP to accelerate the structure degradation.[13] This degradation channel may occur both in presence[14] or absence of light[15] due to the enhanced oxygen electron affinity effect induced by highly-polarized water molecules on the BP surface, but requires both oxygen and water. In order to stabilize BP several passivation strategies involving encapsulation and both non-covalent and covalent surface functionalization have been developed.[16–23] Moreover, very recently a promising coverage procedure employing ionic liquids has been developed. The corresponding protected BP layers were stable for months and the favourable electronic properties could be conserved.[24–26] What the delamination and exfoliation of crystalline BP is concerned, the efficient generation of single layers remains – in contrast to graphite/graphene technology – a very challenging task. For example, the monolayer yield in micromechanical exfoliation is rather limited and the entire process is difficult to control. On the other hand, liquid phase exfoliation using compatible organic solvents leads to some monolayer formation but allows only to the identification of very small flakes with lateral dimensions of less than 500 nm.[27–29] This poses severe limitations to possible applications, *e.g.* in electronic devices. Several efforts have been developed in order to fabricate very thin BP flakes with large lateral dimensions, this includes plasma and laser oxidation processes as well as scanning probe nanolithography and water rinsing.[30–33] In any case, the access to the single layer regime and the control of the number of layers remains a challenge.

Herein we introduce a very simple and highly efficient protocol for the controlled formation of single layer BP. The key point is aqueous step-by-step processing causing continuous layer thinning by oxidative degradation of the outer layers (**Figure 1**). Moreover, we demonstrate that this iterative degradation process can be stopped at will using non-covalent passivation approaches.[17,34] This was accomplished either by surface passivation with perylenes causing effective protection against photo-oxidation, or by protective ionic liquids[24–26], which effectively encapsulate the flakes. Moreover, we demonstrate that the thinning by washing is an appropriate route for the preparation of field effect transistors (FET) exhibiting comparable performance with respect to the pristine BP.

**Results and Discussion**

The oxidative degradation of BP can easily be followed for example by atomic force microscopy (AFM) under ambient conditions with the formation of droplets on the flake surface.[24,35] This is a consequence of the generation of hydrophilic $P_xO_y$ species such as phosphorous acid. These droplets grow continuously until a final coalescence with a dramatic increase in volume is observed. We have now discovered that these droplets can be removed by simple water rinsing in the same way that was previously described for BP-based FET devices. This is a consequence of the high water solubility of the generated phosphorous oxides, leaving a fresh BP surface behind. For the systematic and quantitative monitoring of the decomposition behaviour of thin layer BP, we prepared BP nanosheets of different thicknesses by micromechanical exfoliation of BP using a Scotch Tape. We optimized this procedure by working under the inert conditions of an argon-filled glovebox (<0.1 ppm of $H_2O$ and $O_2$) and starting from a finely grinded sample. The flakes were extensively characterized by scanning Raman microscopy (SRM) and AFM, following our previously published procedure.[6,24] The pristine flakes were characterized after exfoliation, and were subsequently exposed to controlled ambient conditions (average relative humidity: 25%, temperature: 21±1 °C) for several days, leading to BP oxidation. The degradation process was followed by a time-lapse AFM video showing how a BP flake of *ca*. 5 nm is progressively degraded and completely destroyed after 12 hours. (Supporting Information SI I). In order to reduce the thickness, firstly the flakes were rinsed with doubly distilled water after the droplets coalesced on the surface, leading to a new fresh BP surface, this process can be iteratively repeated following the expected layer-by-layer thinning mechanism[35] and showing a progressive and homogeneous nanometric thinning of the flakes in the AFM topography images. An estimation of the reduced thickness of a given flake can be drawn from statistical AFM and Raman analysis of measurements performed after different time spans of controlled oxidation and rinsing, keeping constant the relative humidity, illumination and temperature (Figure 2). Concretely, for the Raman spectroscopy and AFM analysis, the micromechanically exfoliated BP flakes were deposited on 300 nm thick $SiO_2$/Si substrates. Their thickness was determined either by AFM topography images or by monitoring the normalized Raman silicon intensity attenuation by using scanning Raman microscopy (SRM).[24]

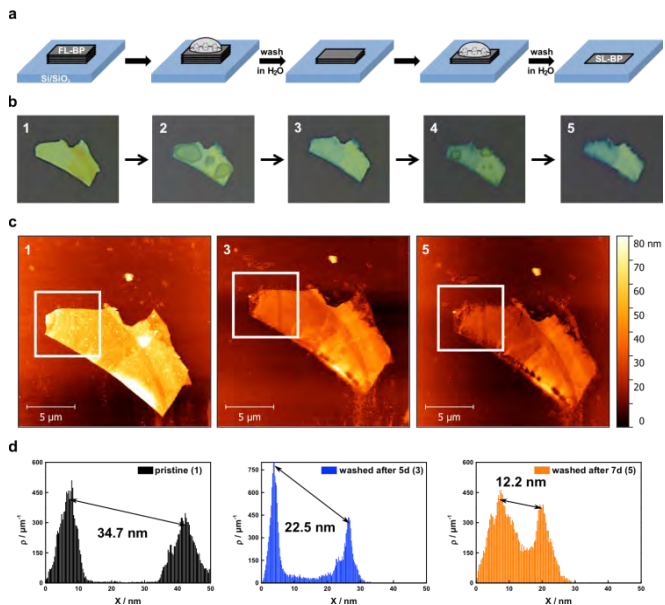

**Figure 1.** Thinning of BP by rinsing with DI water. a) Scheme illustrating the used concept for the layer-by-layer thinning of micromechanically exfoliated BP flakes on Si/SiO$_2$-substrates. During oxidation P$_x$O$_y$-species from (25% relative humidity, 21±1 ºC), which coalesce over time before they can be rinsed with DI water (second step) leading again to a flat BP surface. This process can be repeated several times in order to get the desired thickness. b) Sequence of optical images of a BP flake showing its pristine (1), oxidized after 5 days (2) and the washed (3) form with DI water. Further oxidation for 2 days (4) and subsequent rinsing with DI water (5) complete the series. The color change from yellow to blueish of the flake already indicates a thinning of the BP flake. c) AFM images corresponding to the pristine BP flake (1) as well as after each washing procedure (3&5). d) Statistical AFM evaluation recorded in the white square of the corresponding AFM image above visualizing the difference in height between the underlying substrate and the BP flake which clearly confirming the thinning effect. Note that the first peak in the topography AFM statistics histograms accounts for the substrate, and the second peak to the flake.

A thickness reduction of about 1.5 nm was determined after one day of degradation. After eight days the reduction was more than 14 nm (Figure SI 2). Interestingly, according to AFM analysis, the surface roughness decreases after water rinsing –rough mean square (RMS) values decrease from 1.4 nm (pristine) or from 4.9 nm (oxidized) to 1.1 nm– indicative of a smoother surface (Figure SI 3). Moreover, we investigated the influence of the water rinsing frequency on the degradation kinetics. Firstly, we carried out successive washing steps after every 96 hours, and compared with a day-by-day washing using statistical evaluation of AFM images (Figure SI 4&5). Remarkably, the degradation was slower for the later procedure. This is probably due to the generation of a pristine hydrophobic BP surface, which requires further oxidation (either by O$_2$[9] or by O$_2$ / H$_2$O superclusters[15]) to generate a hydrophilic surface allowing for the physisorption of water, which finally interacts with the previously generated P$_x$O$_y$ leading to phosphorous acid and related species. Therefore, for a precise control over the thinning process, short periodical washing steps are favourable. Notably, we have observed that by increasing the relative humidity to values of *ca*. 60% this degradation process is dramatically accelerated, highlighting the critical role of water in BP degradation kinetics, being more difficult to control the thinning process. In turn, by using our experimental conditions at a controlled humidity of *ca*. 25 %, it is possible to reach an unprecedented control over the thickness of BP flakes, as reflected in the calibration curve obtained for several flakes (Figure SI 2). Indeed, we were able to repeatedly reach a minimum AFM apparent thickness of *ca*. 2.4 nm for several flakes (SI 6-8), which after further degradation were completely degraded, suggesting their monolayer nature. It is worth to mention here that the apparent AFM heights of nanosheets obtained by LPE can be overestimated because of the presence of residual solvent[27,36] as well as contributions from effects such as capillary and adhesion forces, as reported elsewhere.[37,38] In order to elucidate whether we reached the monolayer limit or not, a control experiment measuring monolayer and bilayer graphene flakes prepared by micromechanical exfoliation was performed, obtaining an average thickness of *ca*. 2 nm, thus corroborating the monolayer character of our BP nanosheets (Supporting Information SI 9). Moreover, SRM studies revealed a remaining silicon attenuation of more than 80 %, which perfectly fits to that expected for a monolayer (Figure 2&SI 7).[24]

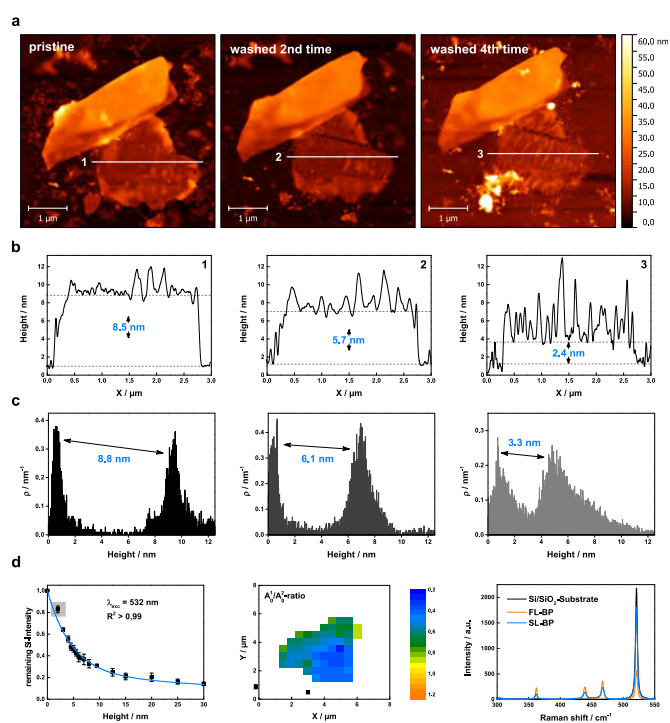

**Figure 2.** Thinning down to the monolayer limit. a) AFM images of a micromechanically exfoliated BP flake, which has been rinsed up to 4 times with DI water in order to reach the monolayer limit. b) Corresponding AFM height profiles along lines 1-3 demonstrating the reduced thickness of the flake. c) Statistical AFM evaluation recorded in the monolayer region of the BP flake which relate nicely to the AFM height profiles. d) Raman characterization of the BP flake: (left). The remaining Si attenuation curve was extended by the point at 2.4 nm. Single layered BP has a remaining Si intensity higher than 80 % in comparison to the pure substrate which represents a nice estimation tool. (Middle) Corresponding Raman mapping of the $A^1_g/A^2_g$-ratio of the BP flake after 4 times of washing with DI water. (right) Mean Raman spectra of the pure substrate, the monolayer region of the BP flake (blue) and the thicker part of the BP flake (orange) highlighting the difference in the remaining SI attenuation.

This top-down chemical methodology can be considered as a new and flexible single-flake exfoliation technique able to provide BP nanosheets with a desired thickness. In order to make it even more tractable it would be desirable to inhibit subsequent degradation step when a given thickness is reached. This would open exciting perspectives for manufacturing of devices. To explore this possibility, we took advantage of two efficient passivation routes previously developed in our group: i) the non-covalent functionalization with tailor-made perylene diimides (PDI)[17], and ii) the passivation using ionic liquids (concretely BMIM-BF$_4$).[24]

To assure comparability, we have prepared flakes of 4 ± 0.5 nm, and then treated them using both passivation approaches. First of all, we investigated the single-flake passivation using a PDI chromophore by dip-coating a thinned BP flake of *ca*. 4.4 nm in a 10$^{-5}$ M solution of the PDI in THF in an argon-filled glove box. The resulting functionalization process was confirmed by statistical Raman spectroscopy (SRS), which shows the dramatic quenching of the PDI fluorescence, with the exclusive PDI functionalization of the BP flakes (Figure 3). We have determined the thickness of the organic layer covering the flakes to be of *ca*. 1.2 nm (Figure SI 10 & 11). A value slightly lower to that theoretically expected for about 6 PDI monolayers, considering the threshold for a complete protection against oxygen and water penetration.[34] Furthermore, to ensure the successful formation of BP-PDI hybrids, XPS measurements were performed. Figure SI 12, shows P 2p core-level photoemission spectra of BP, which exhibits a well resolved single doublet (2p$_{3/2}$ and 2p$_{1/2}$) characteristic of crystalline BP. Moreover, the C 1s and the N 1s component spectrum indicate the presence of carbon and nitrogen atoms corresponding to the conjugated π-system of the PDI and the allylic chains in the peri- position. However, we observed that the PDI-protected flake could survive for about one week. After that, slow degradation takes place. This is probably due to the presence of pinholes and a non-homogeneous coverage of the flake.[39] On the other hand, however, but interestingly, we have observed high stability of flakes under electron-beam irradiation or laser irradiation, thus suggesting a protective effect against photooxidation, a major pathway for BP degradation, as reported elsewhere.[14,40] Note that the extinction spectra of the two samples are very similar (Figure SI 13). Hence, the light-matter interaction (absorption and scattering) in both samples is dominated by BP. The protective action of PDI can, therefore, not consist in protecting BP from light, but rather in protecting it against water and oxygen. Additionally, PDI may also assume a protective role in the photoexcitation dynamics itself, by scavenging photoexcited electrons from the conduction band of BP, thus inhibiting the oxidation. The strong quenching of the PDI fluorescence upon contact with BP suggests that a photoinduced charge or energy transfer indeed happens between PDI and BP.

After the treatment with PDI of the washed BP, the $A^1_g/A^2_g$-ratio of the flakes show no difference, proving the protection of the material against oxidation within a week (Figure SI 14). To study the protective effect of PDI in more detail, we performed for the very first time femtosecond (fs) pump-probe spectroscopy on pure BP and PDI-BP encapsulated in PMMA as an optically inert transparent matrix. The PDI-BP samples were prepared by using the liquid phase exfoliation as previously reported.[13] Sub-100 fs pump pulses of 400 nm wavelength create a population of photoexcited electrons and holes, whose relaxation dynamics is probed *via* monitoring the relative change ΔT/T of the transmission of the broadband probe pulses as a function of the time delay between pump and probe. Contour plots of ΔT/T for BP and PDI-BP are shown in Figures SI 15 and 16. Initially upon photoexcitation, certain absorptive transitions are bleached *via* photoinduced depopulation of their initial and population of their final levels, resulting in positive ΔT/T peaks (increased transmission). The weak positive signal present at 600 nm might be an indicator of a non-negligible population of monolayers as monolayer photoluminescence has been reported on that spectral region. Within a few hundred fs, the spectra of both samples are dominated by a broad photoinduced absorption (reduced transmission) signal resulting from enhanced transitions from photoexcited states to energetically higher levels. Photoexcitation creates a population of hot electron-hole pairs, which relax towards the various band edges within a few hundred fs, as observed previously in BP[41] and other layered two-dimensional semiconductors such as $MoS_2$[42] and $WS_2$.[43] Like the extinction spectrum, also ΔT/T of PDI-BP is dominated by the BP contribution. The small differences in the spectra and dynamics are not sufficient to conclude whether there is an electron transfer from BP to PDI. This would require a dedicated study with a PDI content that largely exceeds what is necessary to protect BP.

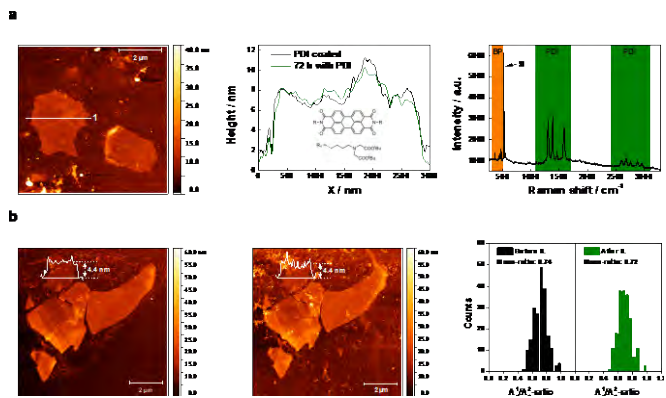

**Figure 3.** Passivation strategies. a) (left)AFM image of micromechanically exfoliated BP covered with PDI. Before the coverage with PDI, the BP has been rinsed with DI water. (middle) AFM height profiles along line 1 after the coverage with PDI and 72 hours later. The nearly same shape of the illustrates the passivation effect of the PDI, temporarily preventing any oxidation phenomena. The inset shows the molecular structure of the used PDI: N-Bis-(tert-butyl-(2,2'-aminobutylazanediyl)-diacetate)-3,4,9,10 perylene diimide. (right) Mean Raman Spectra of the non-covalently functionalized BP flake with PDI showing both the characteristic peaks of the BP as well as the fine structure of PDI peaks. b) (left) AFM image of micromechanically exfoliated BP which has been treated several times with DI water before IL was deposited to prevent oxidation. (middle) AFM image of the corresponding flakes after they had been covered for 10 days by IL. (right) Comparison of the $A^1_g/A^2_g$-ratio of the BP flakes before and after treatment with IL showing nearly no difference.

The intense pump light of 5.1 W/cm$^2$ at 400 nm makes the pump-probe experiment an excellent tool to directly monitor the photooxidation of BP. Figures SI 17-19 show the ΔT/T signal at a selected wavelength for consecutive scans of 20 min duration each. While for the two PDI-BP samples the signal changes very little from one scan to the next (especially after the first three scans), for pure BP the signal decreases strongly during the measurement. The decrease is permanent and does not recover after waiting for several minutes, while repeating the measurement at a fresh spot yields a similar signal to the first scan reported, showing that BP is indeed photodegraded by the pump laser and that PDI protects it against such degradation.

On the other hand, the use of ionic liquids (ILs) is a very promising strategy due to their outstanding ability to create compact passivation layers on the BP surface avoiding degradation for months,[24] probably due to the ability of imidazole ILs for trapping several reactive oxygen species (ROS) generated by photo-oxidation.[25,26] Along these lines, we prepared a sample using our oxidative layer-by-layer thinning method to reach a reduced thickness of *ca*. 4.4 nm (Figure SI 20), before depositing a thin layer of BMIM-BF$_4$. We monitor the flake by SRM after 10 days of passivation, a period of time much longer than that expected for the degradation of flake with this thickness and lateral dimensions. Figure 3 shows the $A^1_g/A^2_g$ histograms of the selected flakes before and after 10 days of the IL deposition under environmental conditions exhibiting a negligible average decrease of *ca.* 0.02, indicative of an effective passivation. Further analysis can be found in the supporting information Figure SI 21.

Last but not least, as a proof of concept and in order to evaluate the potential of our thinning approach for applications in electronic devices, we have prepared a field effect transistor with a BP flake reduced until about 10 nm. It is worth to mention that, for most of the applications flakes in the range around 10 nm are of fundamental interest.[3] The initial height of the flake was on average 14 nm (Figure SI 22). After the thinning procedure the flake did not show any apparent large-scale damage on the surface. Thus, immediately after thinning we have contacted the flakes using a FET configuration, as can be clearly observed in optical-, scanning-force and scanning-electron microscopies (Fig. 4 and Figure SI 23, respectively). The variation in height was recovered after the thinning procedure. In detail, the height determined by scanning-force microscopy ranged about 7 to 10 nm on one side, and 13 to 17 nm on the other side of the sample (SI 22 & 23). On the basis of these results, the mobility and carrier-density was determined for each side of the sample separately. Our electrical measurements revealed for both sides *p*-type-conduction with a carrier density of the order of $10^{19}$ cm$^{-3}$. Both, the *p*-type conduction and the carrier-density range are in agreement with previous reports for as-synthesized as well as mechanically exfoliated black phosphorus as we used it in our study.[44–46] For the mobilities, we found 74 ± 2 cm$^2$/Vs (7 to 10 nm height) and 65 ± 5 cm$^2$/Vs (13 to 17 nm height). These values are within the range of experimental mobilities reported in literature for similar samples, including those submitted to chemical etching using DI water or scanning probe nanopatterning.[2,33,47] More precisely, the mobility value of the thinned BP is slightly lower than for a pristine micromechanically exfoliated BP flake of similar thickness, but its carrier density and therefore conductivity (340 S/cm) are enhanced. These results demonstrate that the thinning procedure yields flakes that are not electrically compromised. As a consequence, the way for the use of aqueous controlled oxidation for the preparation of electronic devices with controlled thickness is paved.

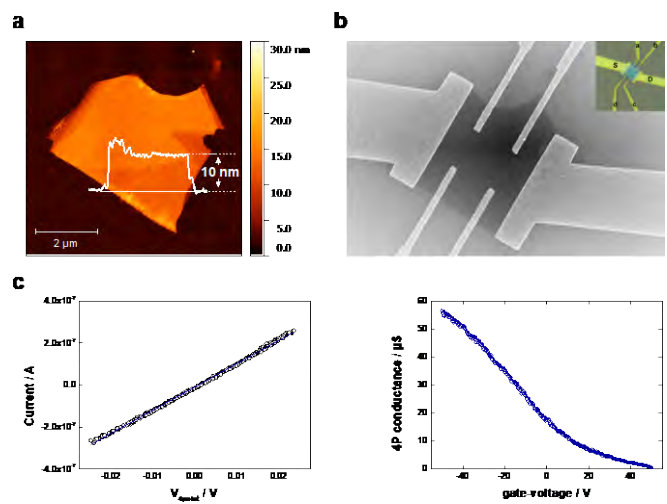

**Figure 4.** Field effect transistor (FET). a) AFM image of a micromechanically exfoliated BP flake which has been rinsed with DI water prior to electrical measurements. SEM image of the rinsed BP flake which has been contacted with a mixture of Ti/Au contacts. The inset shows an Optical image of the contacted flake with a labelling of each contact. c) Electrical characterization: (left) Recorded I-V curve over a span of 50 mV. (right) 4P conductance measurement.

## Conclusions

In summary, we have developed an unprecedented top-down strategy for reducing the thickness of BP flakes at will by controlling the oxidation of the flakes and removing the oxidized phosphorous species (*e.g.* phosphorous acid) with distilled water. As a result we can initiate a successive layer-by-layer thinning process. The iterative thinning procedure can be terminated at will either by using non-covalent functionalization with perylene diimide molecules, which act as a protecting barrier against photooxidation as demonstrated by femtosecond transient spectroscopy, or by using BMIM-BF$_4$ ionic liquid as a passivation layer. The suitability of this thinning procedure is revealed through the preparation of FET devices, showing that the electronic properties of flakes are not compromised. This work provides fundamental insights into a straightforward procedure for controlling the thickness of BP flakes keeping intact its electronic properties, and holds great promise in the future development of BP-based opto-electronic devices.

## Experimental Section

*Materials and exfoliation process:* Throughout all experiments, BP with purity higher than 99.999% (Smart Elements) was used. BP nanosheets were produced by mechanical exfoliation with a commercially available Scotch Tape (3M). Afterwards, the FL-BP flakes were transferred onto Si/SiO$_2$-substrates (300 nm oxide layer). The exfoliation was performed in an argon filled LABmasterpro sp glove box (MBraun) equipped with a gas purifier and solvent vapour removal unit (oxygen and water content lower than 0.1 ppm).

*Solvent purification:* Anhydrous, 99,9% purity solvents (THF, NMP and 1-butyl-3-methylimidazolium tetrafluoroborate (BMIM-BF$_4$)) were purchased from Sigma-Aldrich. The BMIM-BF$_4$ was degassed under vacuum during 3 days prior to the experiments. THF and NMP were first dried over molecular sieve (3 Å) for at least 3 days to remove dissolved water. The water content was determined by using the "Karl Fischer" method, obtaining values lower than 5 ppm. Additionally, both solvents were degassed by iterative pump freezing (a minimum of 7 cycles) in liquid nitrogen to remove oxygen, before introducing them into the glove box.

*On-surface non-covalent functionalization with N-Bis-(tert-butyl-(2,2'-aminobutylazanediyl)-diacetate)-3,4,9,10 perylene diimide (PDI):* The exfoliated few-layer BP samples were functionalized by dip-casting them for <2 s into a tailormade EDTA-PDI (EDTA=ethylenediaminetetraacetic acid) derivative dissolved in THF (c = 10$^{-5}$ M).[17,48] Residual PDI was removed by drop casting isopropanol during spin coating of the samples.

*Characterization:* Immediately after the removal from the inert atmosphere, images of FL-BP flakes were recorded under an optical microscope (Zeiss Axio Imager M1m), using different objectives enabling their re-localization in Raman and AFM measurements.

Raman spectra were acquired on a LabRam HR Evolution confocal Raman microscope (Horiba) equipped with an automated XYZ table using 0.80 NA objectives. All measurements were conducted using an excitation wavelength of 532 nm, with an acquisition time of 2 s and a grating of 1800 grooves/mm. To minimize the photo-induced laser oxidation of the samples, the laser intensity was kept at 5 % (0.88 mW). The step sizes in the Raman mappings were in the 0.5–1 µm range depending on the experiments. Data processing was performed using Lab Spec 5 as evaluation software. When extracting mean intensities of individual BP Raman modes, it is important to keep each spectral range constant, e.g. from 355–370 cm-1 and from 460–475 cm-1 because otherwise the resulting value of the $A^1_g/A^2_g$-ratio can be slightly influenced.

Atomic force microscopy (AFM) was carried out using a Bruker Dimension Icon microscope in tapping mode. Bruker Scanasyst-Air silicon tips on nitride levers with a spring constant of 0.4 N·m$^{-1}$ were used to obtain images resolved by 512x512 or 1024x1024 pixels.

*X-ray photoelectron spectroscopy (XPS):* XPS was performed on a commercially available Quantera II instrument (Physical Electronics Inc., Chanhassen, MN, USA). The powder samples were mounted to the sample holder on a double-sided adhesive tape. Monochromated Al Kα X-rays (1486.6 eV) from a source operating at 15 kV and 25 W were focused on 100 × 100 µm$^2$ homogeneously covered spots on the samples. The floating samples were neutralized by Ar$^+$ ion and electron bombardment during measurement. Emitted photoelectrons were detected at an analyzer pass energy of 112 eV, 55 eV and 13 eV for detailed elemental spectra of N 1s, C 1s and P 2p, respectively. The binding energy scale was referenced to the adventitious C 1s core-level at 284.8 eV.

*Absorption spectroscopy:* UV/Vis measurements were performed using Perkin Elmer Lambda 1050 spectrometer in extinction, in quartz cuvettes with a path length of 0.4 cm using BMIM-BF4 as reference.

*Pump-probe experiments:* Transient absorption measurements have been performed in a fs Ti:sa laser set up. A pulsed laser with a 250KHz repetition rate, centered at a wavelength of 800nm with a continuous wave power of 800mW is our light source, as the irradiation comes from a pulsed laser of 250KHz repetition rate of 150 fs duration pulses that leads to a 24mW peak power. Laser beam is attenuated and split in pump and probe beam. Pump beam's frequency is doubled in a barium borate crystal, resulting in a pump wavelength of 400nm and a pump power of 0.9 mW. The probe beam pulses are focused in a sapphire plate and supercontinuum white light probe pulses (460–760 nm) are generated. Probe beam continuous wave power is set at 10 µW. Pump and probe beams are focused and overlapped in the sample. Leading to a spot size of 150 µm. Measurements have been done in transmission, pump beam has been blocked and the probe beam is collimated and guided to the detector.

*Electrical measurements:* Flakes were prepared by mechanical exfoliation under inert atmosphere on n++ (As)-doped silicon wafers with 300 nm thermally grown SiO$_2$ on top. Typical lateral flake sizes were a few µm. For electrical contacting standard electron beam-lithography was used. Electrical contacts were defined by evaporation of 5 nm titanium and 40 nm gold on top of the sample. After deposition of the electrodes the samples were covered by PMMA to protect from air. The electrical measurements to determine the four-point conductivity G and from this the field-effect mobility µ were carried out at room-temperature where the doped silicon served as a backgate. The mobility µ was extracted from the linear part of the gate-dependence of G following

$$\mu = \frac{t_{ox}}{\varepsilon_0 \varepsilon_{SiO_2}} \frac{L}{W} \frac{dG}{dV_G}$$

with the thickness $t_{ox}$ of the SiO$_2$, the dielectric constants $\varepsilon_0$ and $\varepsilon_{SiO_2}$, channel length $L$ and width $W$ (any hysteretic behavior was not further pursued as attributed to stem from commonly appearing inhomogeneity at the flake substrate interface in 2D layered materials FETs). The height of flakes was measured by scanning force microscopy.

## Conflicts of interest
There are no conflicts to declare.

## Acknowledgements
A.H. and G.A. acknowledge the European Research Council (ERC Advanced Grant 742145 B-PhosphoChem to A.H., and ERC Starting Grant 2D-PnictoChem 804110 to G.A.) for support. The research leading to these results was partially funded by the European Union Seventh Framework Programme under grant agreement No. 604391 Graphene Flagship. G.A. has received financial support through the Postdoctoral Junior Leader Fellowship Programme from "la Caixa" Banking Foundation. G.A. thanks support by the Deutsche Forschungsgemeinschaft (DFG; FLAG-ERA AB694/2-1), the Generalitat Valenciana (SEJI/2018/034 grant) and the FAU (Emerging Talents Initiative grant #WS16-17_Nat_04). V.K. and A.H. thank the SFB 953 "Synthetic Carbon Allotropes" funded by the DFG for support and the Cluster of Excellence „Engineering of Advanced Materials". Financial support by MINECO through the Excellence Unit María de Maeztu (MDM-2015-0538) is acknowledged.

## Notes and references

# Supporting Information

**Monolayer black phosphorus by sequential wet-chemical surface oxidation**

*By Stefan Wild, Vicent Lloret, Victor Vega-Mayoral, Daniele Vella, Edurne Nuin, Martin Siebert, Maria Koleśnik-Gray, Mario Löffler, Karl J. J. Mayrhofer, Christoph Gadermaier, Vojislav Krstić, Frank Hauke, Gonzalo Abellán\* and Andreas Hirsch\**

S. Wild, V. Lloret, Dr. E. Nuin, Dr. F. Hauke, Dr. G. Abellán, Prof. A. Hirsch
Department of Chemistry and Pharmacy and Joint Institute of Advanced Materials and Processes (ZMP)
Friedrich-Alexander-Universität Erlangen-Nürnberg (FAU)
Nikolaus Fiebiger-Strasse 10, 91058 Erlangen and Dr.-Mack Strasse 81, 90762 Fürth, (Germany).

Dr. V. Vega-Mayoral
CRANN & AMBER Research Centers
Trinity College Dublin, Dublin 2, Ireland & School of Physics
Trinity College Dublin, Dublin 2, Ireland

Dr. V. Vega-Mayoral, Dr Daniele Vella, Prof. C. Gadermaier
Department for Complex Matter
Jozef Stefan Institute
Jamova 39, 1000 Ljubljana, Slovenia & Jozef Stefan International Postgraduate School, Jamova 39, 1000 Ljubljana, Slovenia

Dr. Daniele Vella Department of Physics National University of Singapore, 2 Science Drive 3, Singapore 117542, Singapore.

Dr. E. Nuin, Dr. G. Abellán
Instituto de Ciencia Molecular (ICMol), Universidad de Valencia,
Catedrático José Beltrán 2, 46980, Paterna, Valencia, Spain

M. Siebert, Dr. M. Koleśnik-Gray, Prof. V. Krstić
Department of Physics
Friedrich-Alexander-Universität Erlangen-Nürnberg (FAU)
Staudtstr. 7, 91058 Erlangen, Germany.

M. Löffler, Prof. K. J. J. Mayrhofer
Helmholtz Institute Erlangen-Nürnberg for Renewable Energy (IEK-11),
Forschungszentrum Jülich GmbH,
Egerlandstraße 3, 91058 Erlangen, Germany
Department of Chemical and Biological Engineering,
Friedrich-Alexander-Universität Erlangen-Nürnberg (FAU), Immerwahrstraße 2a,
91058 Erlangen, Germany

Prof. C. Gadermaier
Department of Physics
Politecnico di Milano
Piazza Leonardo da Vinci 32
20133 Milano, Italy

## SI 1:

AFM video showing the complete degradation of a BP flake (see file attached).

## SI 2:

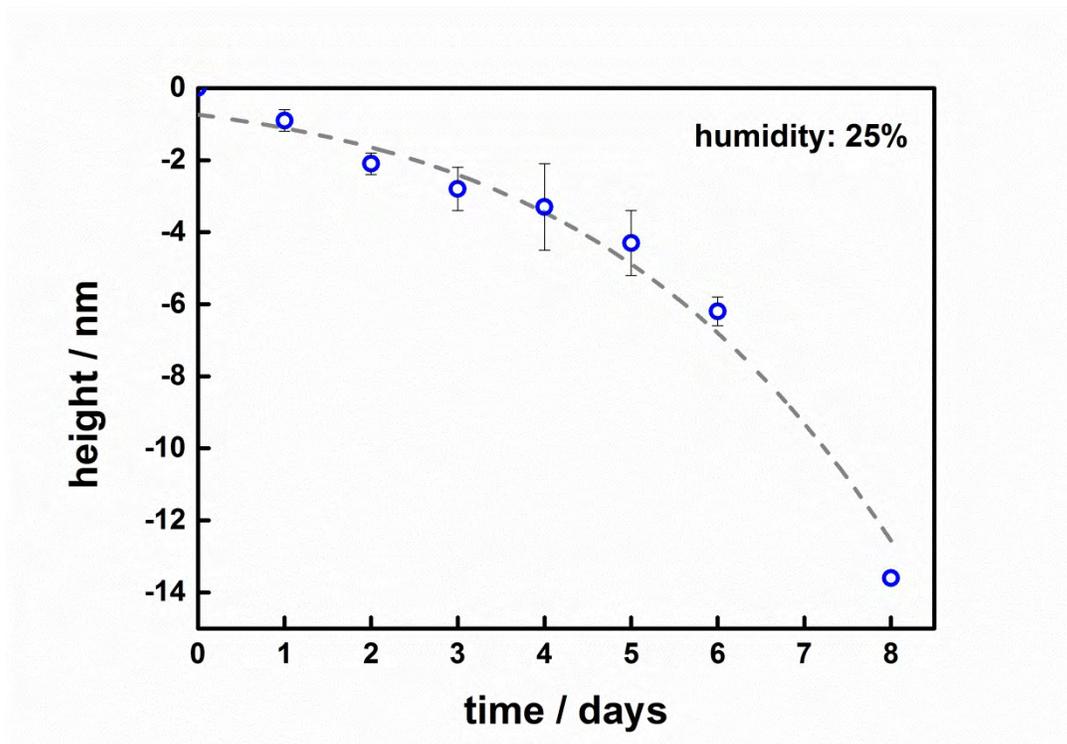

**Figure SI 2:**
Calibration curve showing the thickness of BP flakes versus the time when treated with DI water. It is very important to remark that the humidity highly influences these values, e.g. when reaching humidity values higher than 60 % the reduced thickness can be twice in the same amount of time for this thinning technique.

**SI 3:**

| | $R_{ms}$ / nm (Flake1) | $R_{ms}$ / nm (Flake2) |
|---|---|---|
| pristine | 1.34 | 1.48 |
| 1d oxidized | 5.42 | 5.87 |
| 2d oxidized | 6.10 | 5.37 |
| 3d oxidized | 4.85 | 5.03 |
| washed | 1.09 | 1.12 |
| PDI coated | 1.07 | 1.04 |

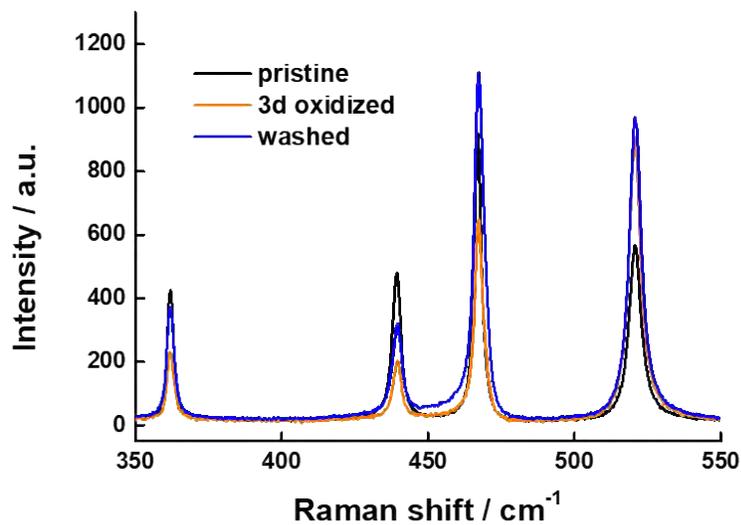

**Figure SI 3:**

(Top): Evolution of the surface roughness of BP flakes during oxidation, after rinsing with DI water and after the non-covalent attachment of PDI. (Bottom): Evolution of the Mean BP Raman intensities: Upon oxidation, BP Raman intensities decrease, but interestingly, when rinsed with DI water, the BP Raman intensity gets recovered which fits perfectly to the recovery of the pristine BP surface.

**SI 4:**

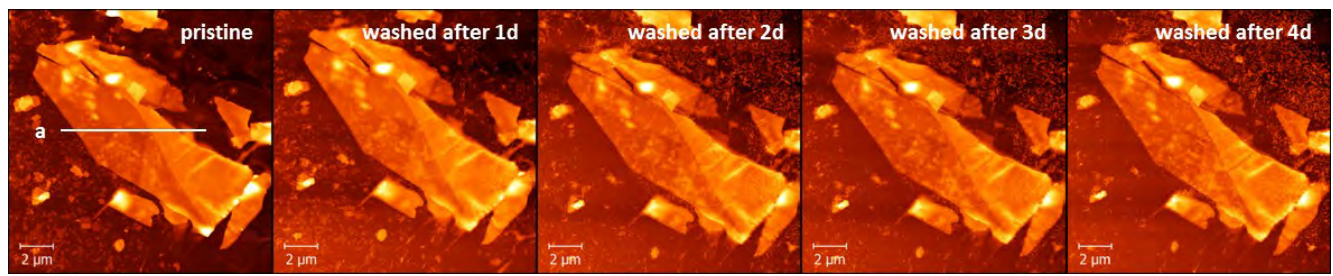

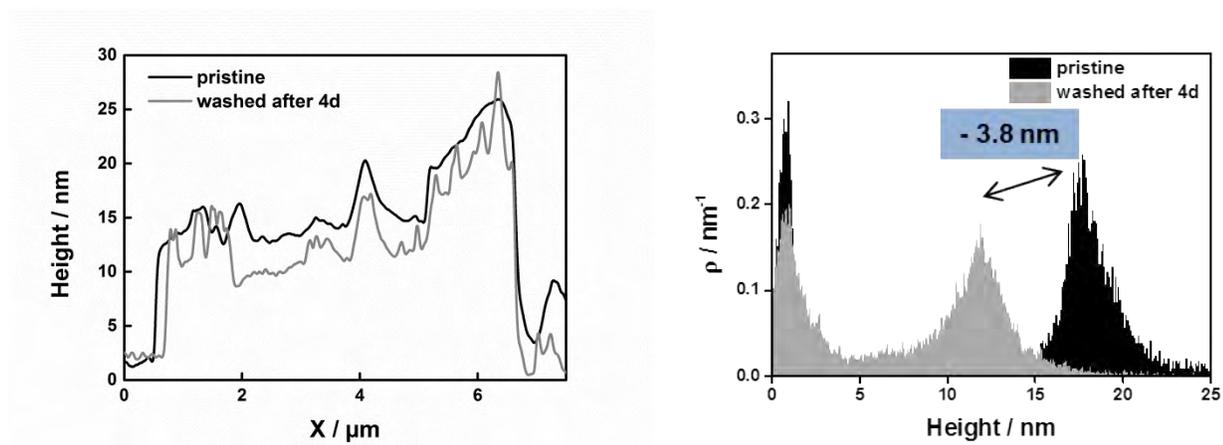

**Figure SI 4:**

(Top): AFM image of mechanically exfoliated FL-BP which has been washed 4 times (daily) with DI water. (Bottom): AFM height profiles along line a (left) and the related statistical AFM evaluation (right) comparing the pristine thickness to the thickness of the flake after 4 washing procedures.

**SI 5:**

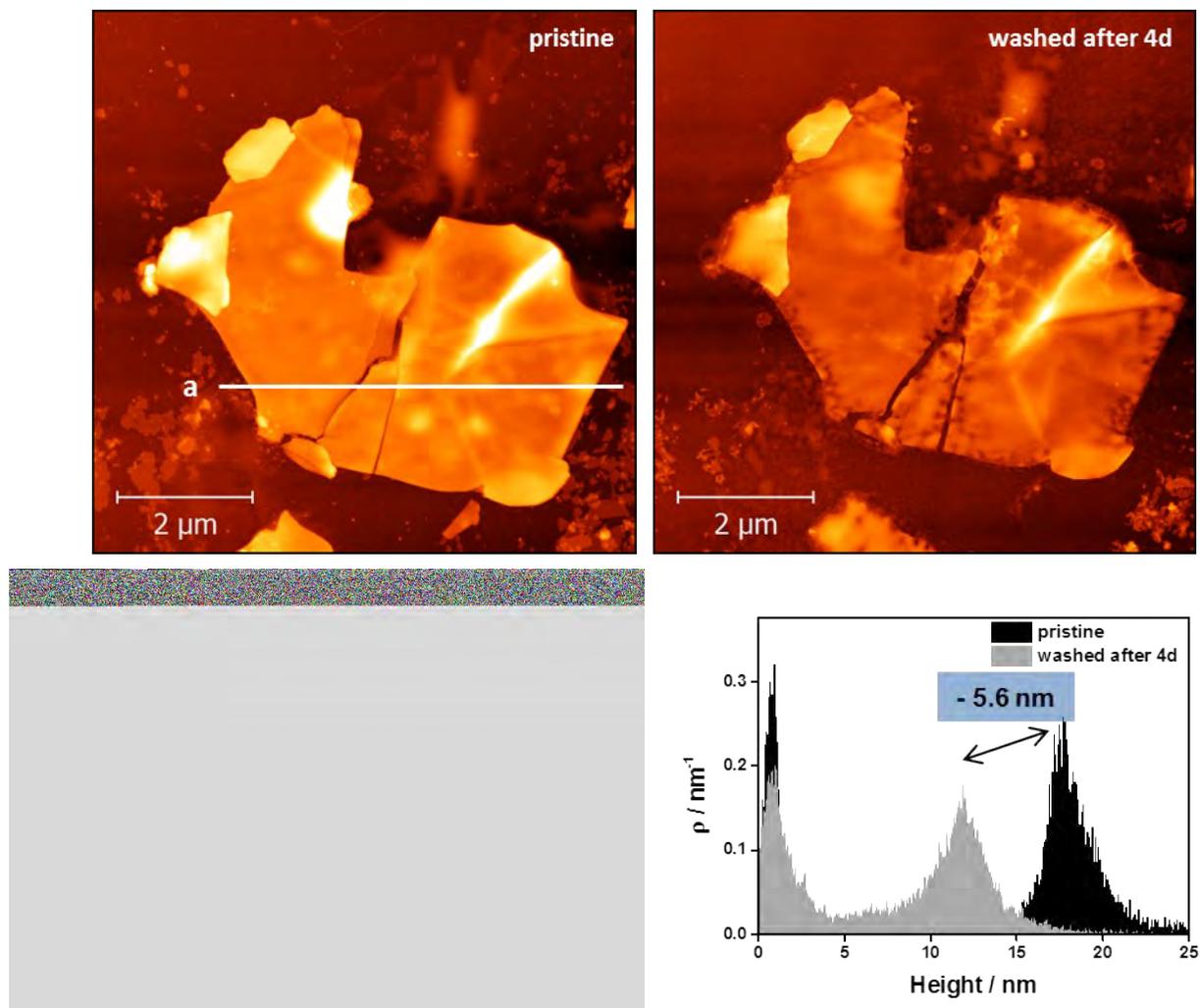

**Figure SI 5:**

(Top): AFM image of mechanically exfoliated FL-BP which has been washed after 4 days with DI water. (Bottom): AFM height profiles along line a (left) and the related statistical AFM evaluation (right) comparing the pristine thickness to the thickness of the flake washed after 4 days. Comparing to the flake shown in figure SI 6 it is worth to remark that washing the BP flake every day results in a slower degradation than washing only one time in the same amount of time.

**SI 6:**

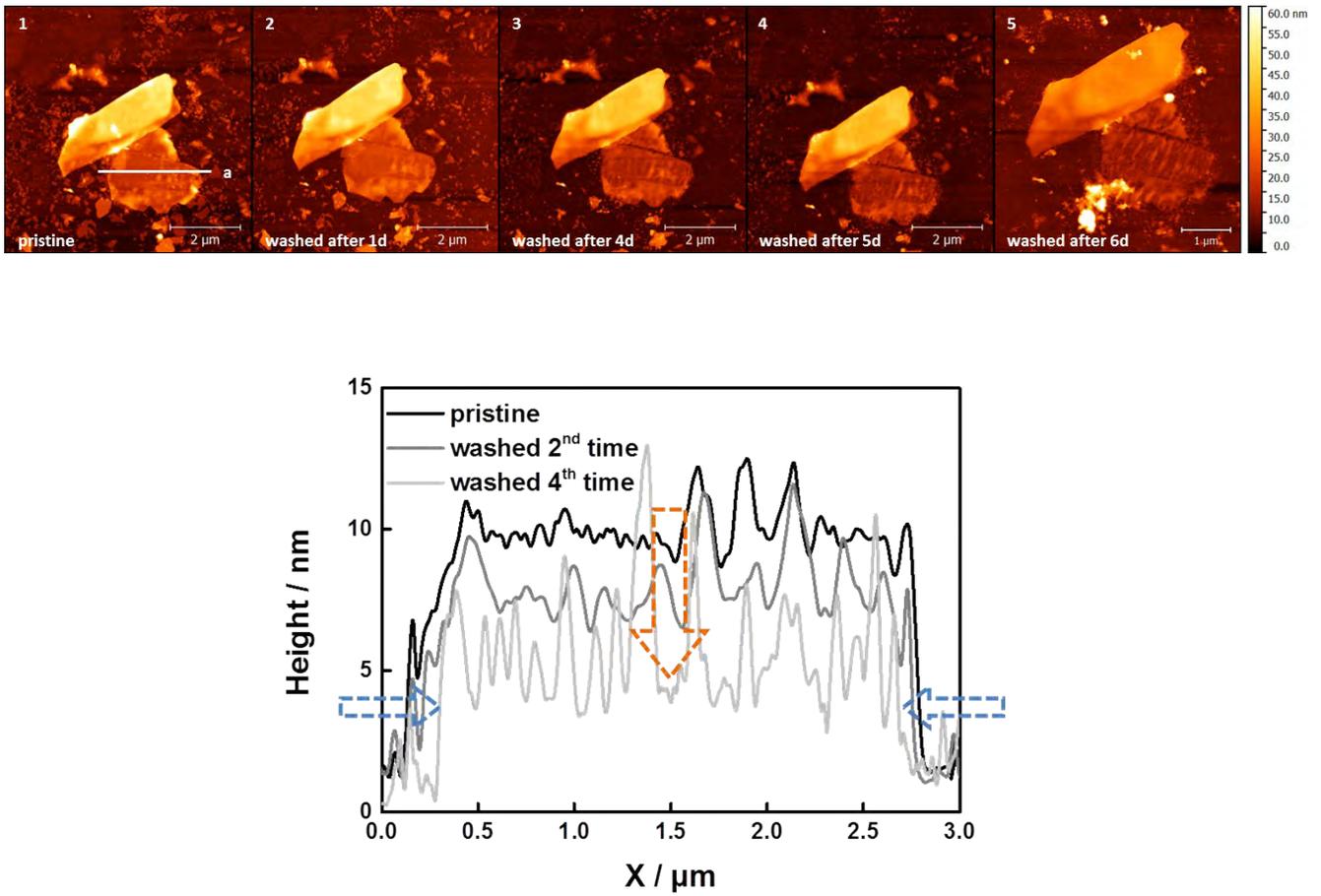

**Figure SI 6:** (Top): Sequence of AFM images depicting a mechanically exfoliated BP flake which has been washed several times with DI water in order to produce SL-BP. (Bottom): Corresponding AFM height profiles along line a. Interestingly, not only a decrease in the thickness of the flake can be observed, but also a reduce in the lateral dimensions of the flake. This confirms on the on the hand that the degradation of black phosphorous follows the layer by layer thinning effect described by Castellanos[1] but on the other hand oxidation of the edges also plays a role.

**SI 7:**

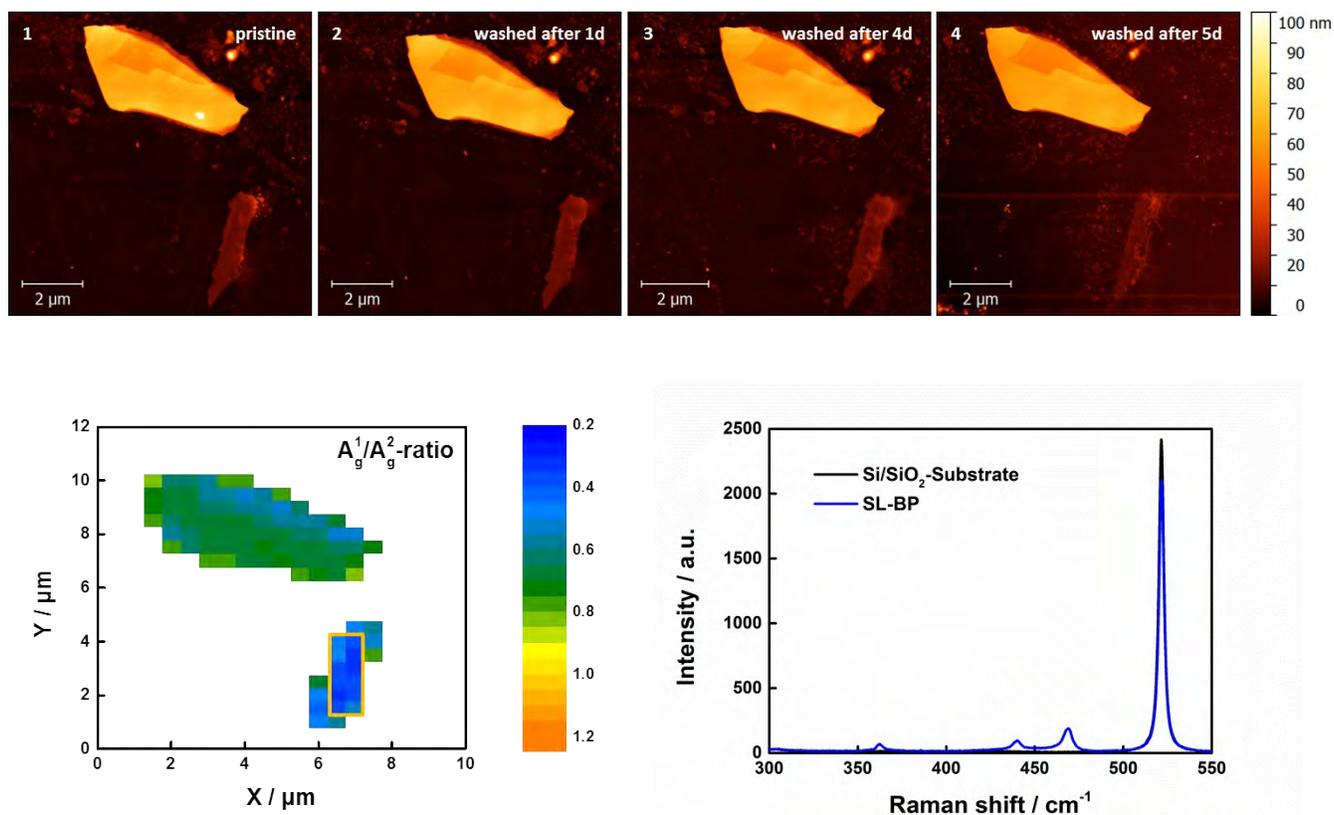

**Figure SI 7:**

(Top): Sequence of AFM images depicting mechanically exfoliated BP flakes which have been washed several times with DI water in order to produce SL-BP. (Bottom-Left): Raman mapping of the $A^1_g/A^2_g$-ratio of the BP flakes which has been conducted after the last washing step. (Bottom-Right): Mean Raman spectrum (orange area) of the thin BP flake in comparison to the Si/SiO$_2$ substrate. The high remaining silicon intensity (>82%) indicates that the flake consist only of one BP layer.

**SI 8:**

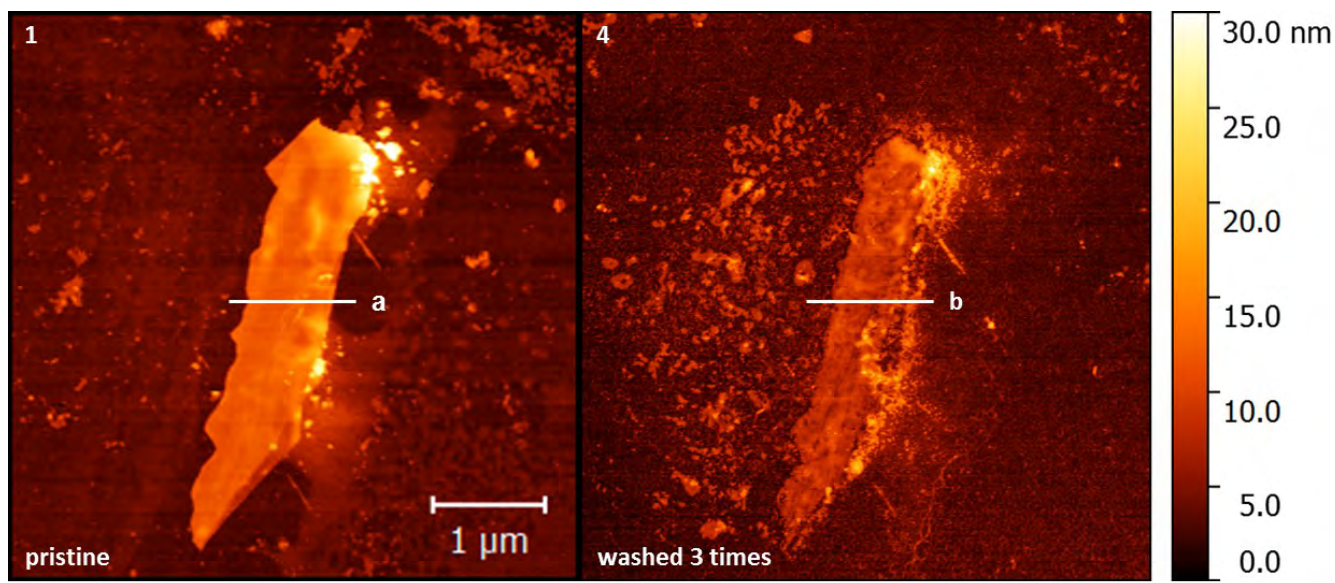

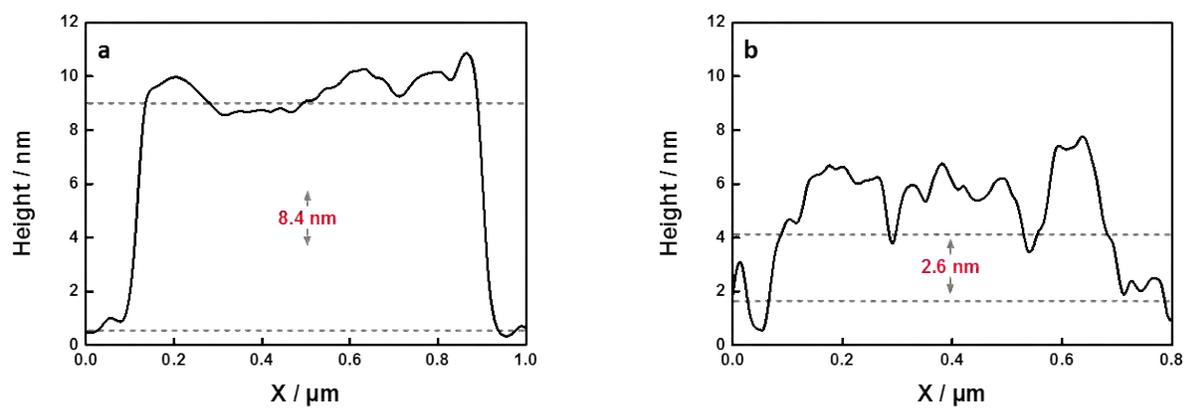

**Figure SI 8:**

(Top): Zoom-in of the AFM images shown in figure SI 3. Comparison of the pristine BP flake and the same flake after it has been dip-casted 3 times in DI-water to reduce the thickness down to SL-BP. (Bottom): Corresponding AFM height profiles taken along line a and b.

**SI 9:**

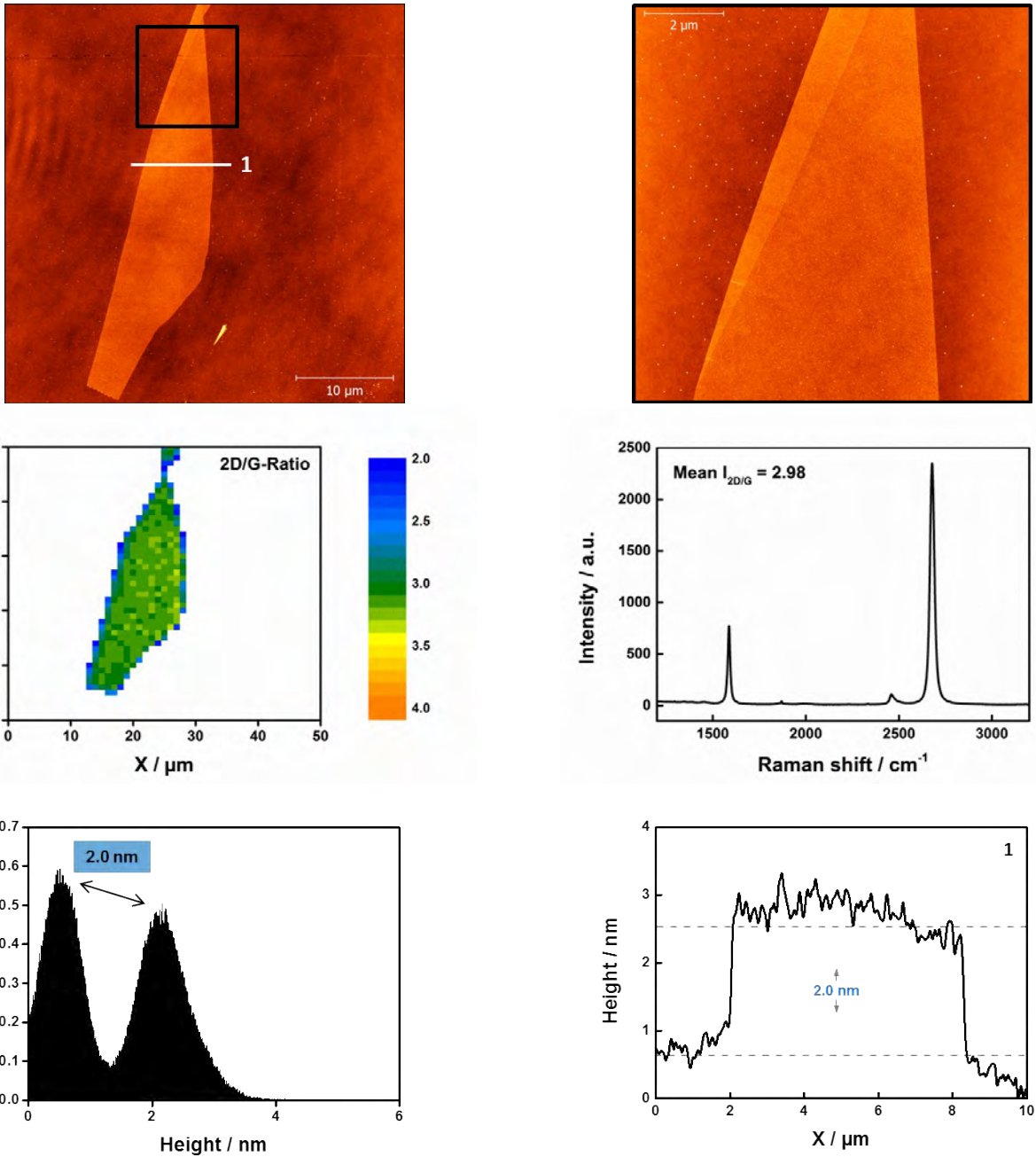

**Figure SI 9:**

(Top): AFM image of a mechanically exfoliated SL-graphene flake (left) and a zoom-in highlighting that a small part of the flake is folded. (Middle): Corresponding Raman mapping of the SL-graphene showing the 2D/G-ratio (left) and the related Mean Raman spectra (right) with a mean $I_{2D/G}$-ratio of 2.98 confirming that SL-graphene is present. (Bottom): Statistical AFM evaluation (left) visualizing the thickness of the SL-graphene flake and the corresponding AFM height profile along line 1.

**SI 10:**

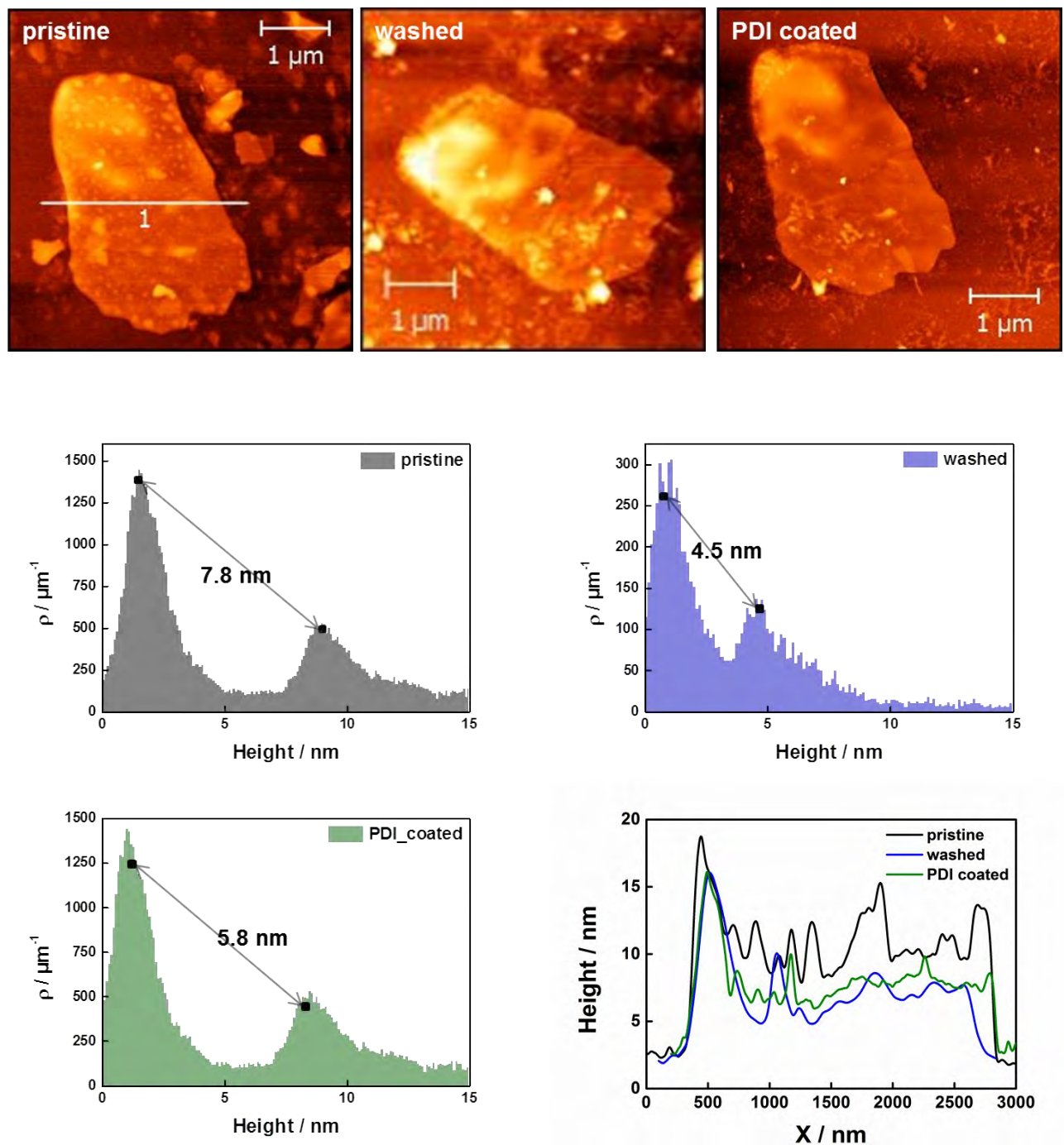

**Figure SI 10:**

(Top): AFM image of mechanically exfoliated FL-BP which has been washed with DI water and afterwards coated with PDI by dip-casting the wafer in a $10^{-5}$M PDI solution. The non-covalently attached PDI temporarily prevents oxidation of the BP. (Middle & Bottom): Statistical AFM evaluation showing in a first step the reduced thickness of the BP flake when washed with DI water and in the second step an increased thickness because of the PDI layer on top of the BP flake. Corresponding AFM height profiles along line 1 confirm this trend.

**SI 11:**

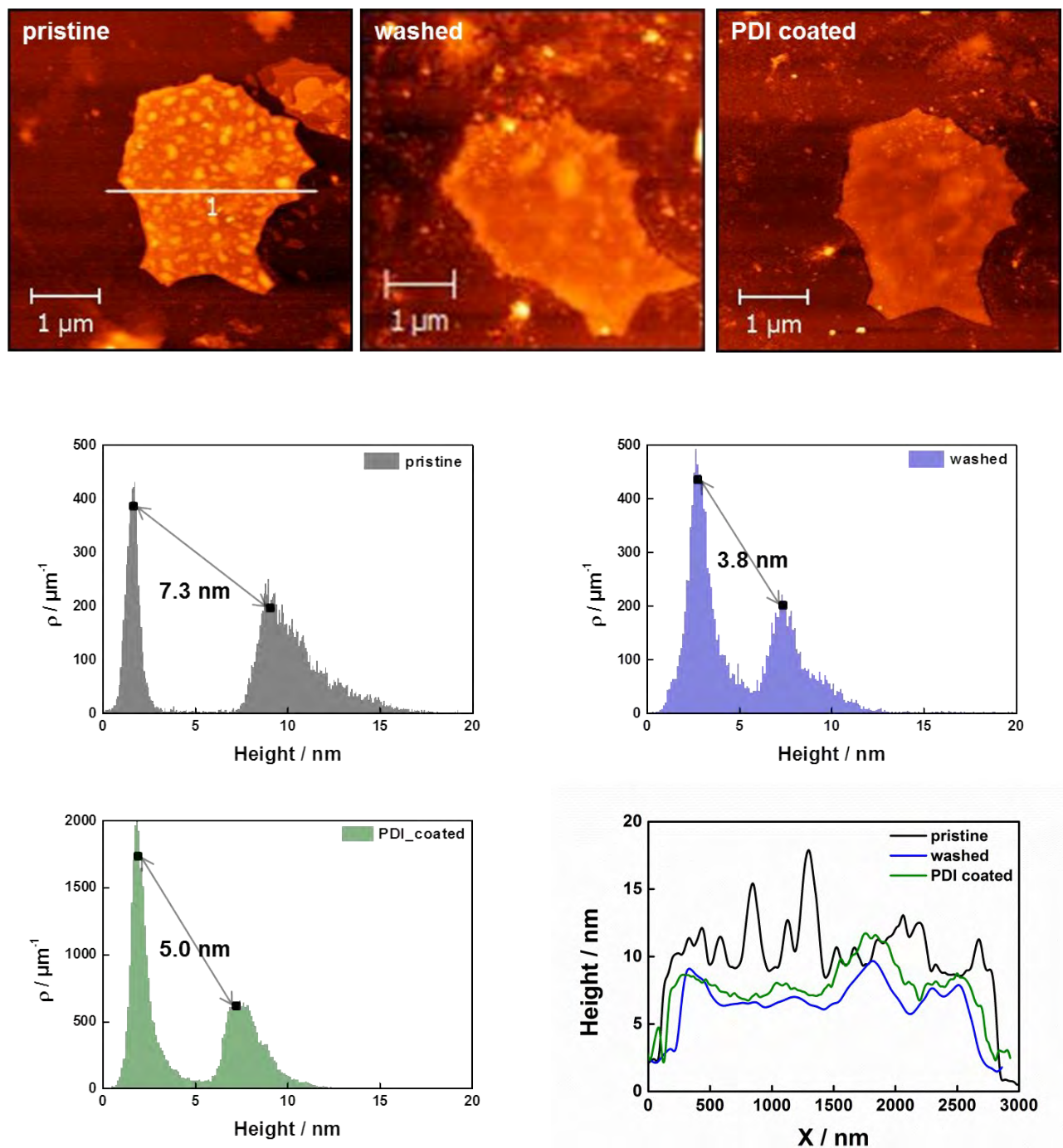

**Figure SI 11:**

(Top): AFM image of mechanically exfoliated FL-BP which has been washed with DI water and afterwards coated with PDI by dip-casting the wafer in a $10^{-5}$M PDI solution. The non-covalently attached PDI temporarily prevents oxidation of the BP. (Middle & Bottom): Statistical AFM evaluation showing in a first step the reduced thickness of the BP flake when washed with DI water and in the second step an increased thickness because of the PDI layer on top of the BP flake. Corresponding AFM height profiles along line 1 confirm this trend.

**SI 12:**

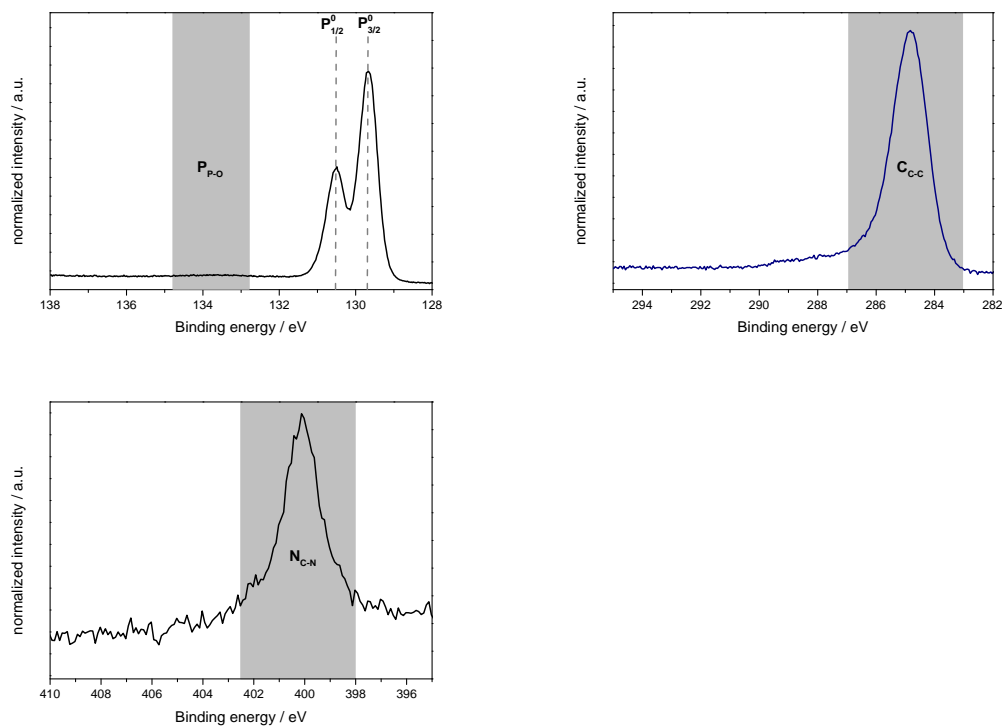

**Figure SI 12A** XPS BP-PDI hybrid showing the regions of P, without any oxidation, and the regions of C and N indicating that the perylene is interacting with BP.

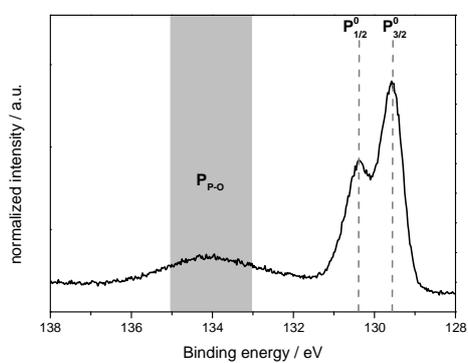

**Figure SI 12B** XPS of pristine BP showing the regions of P, with the typical oxidation P-O bonds.

**SI 13:**

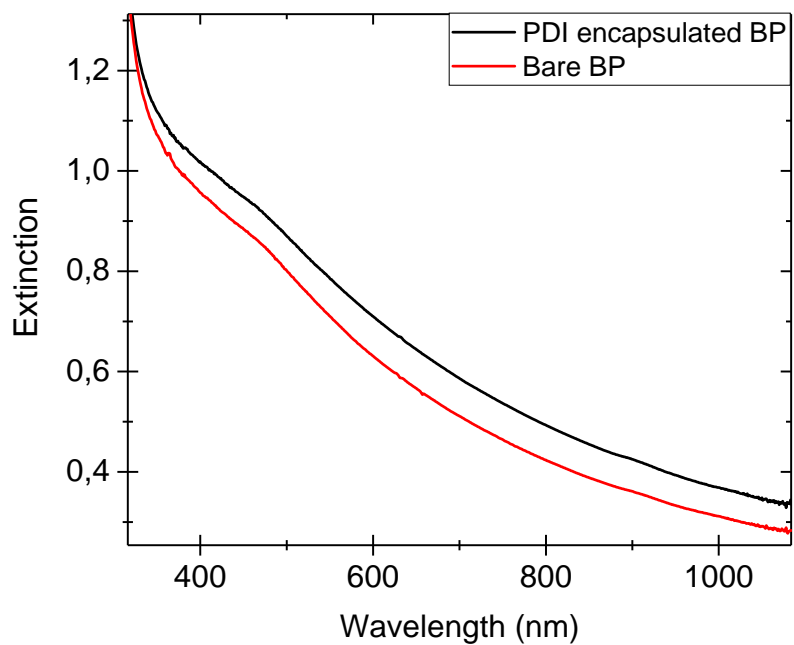

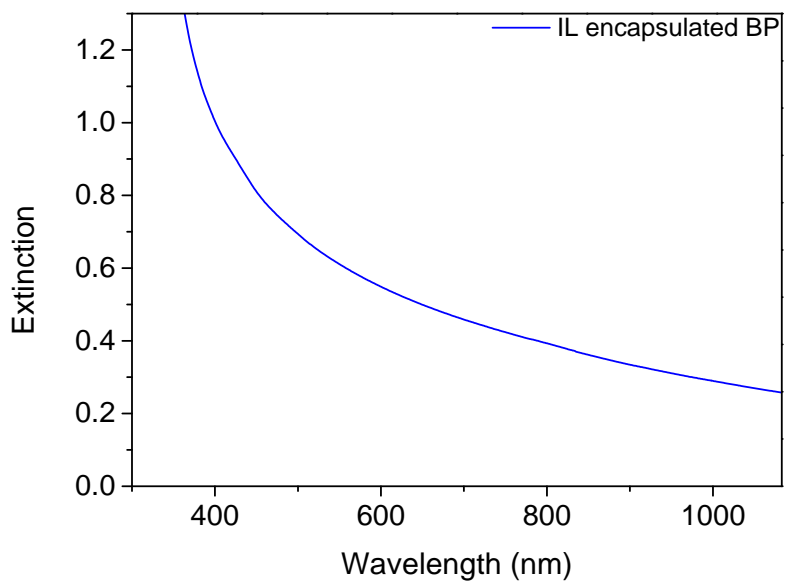

**Figure SI** 13 Extinction spectra of bare BP and BP encapsulated with perylene.

**SI 14:**

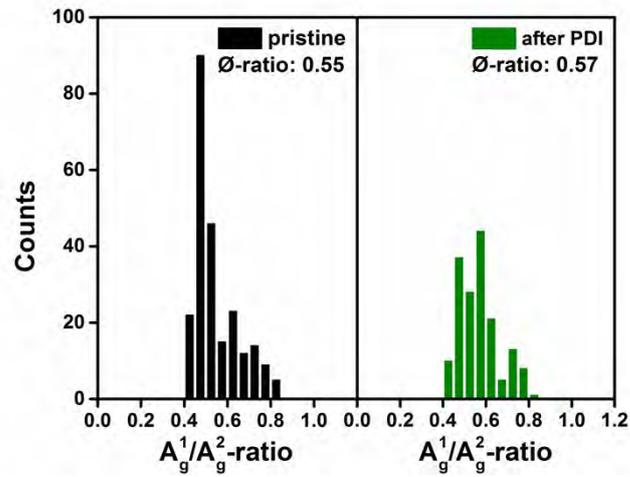

**Figure SI 14** Comparison of the $A^1_g/A^2_g$-ratio of the BP flakes before and after treatment with BP showing nearly no difference.

**SI 15:**

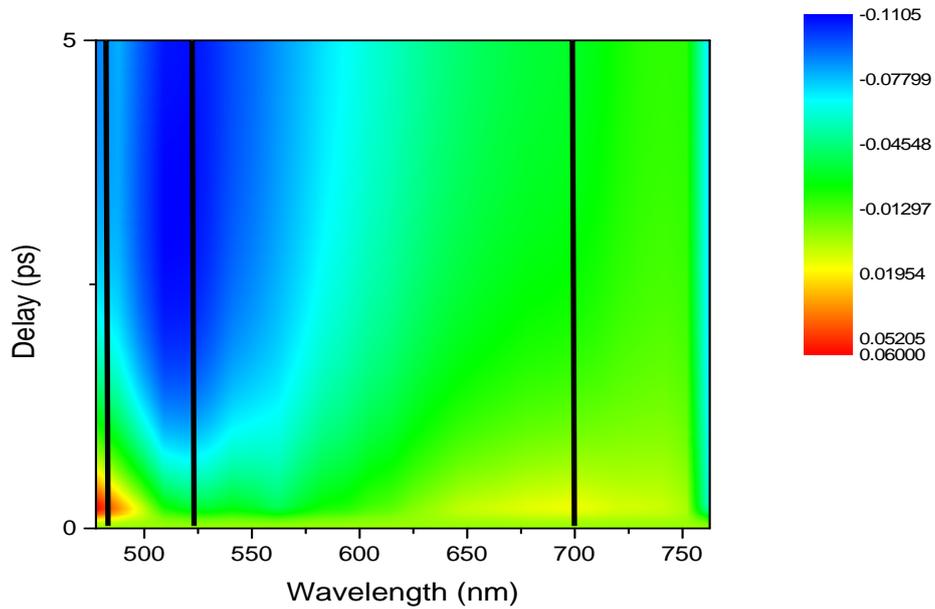

**Figure SI 15** Contour plot of ΔT/T for a pristine exfoliated BP sample encapsulated in PMMA as an optically inert transparent matrix.

**SI 16:**

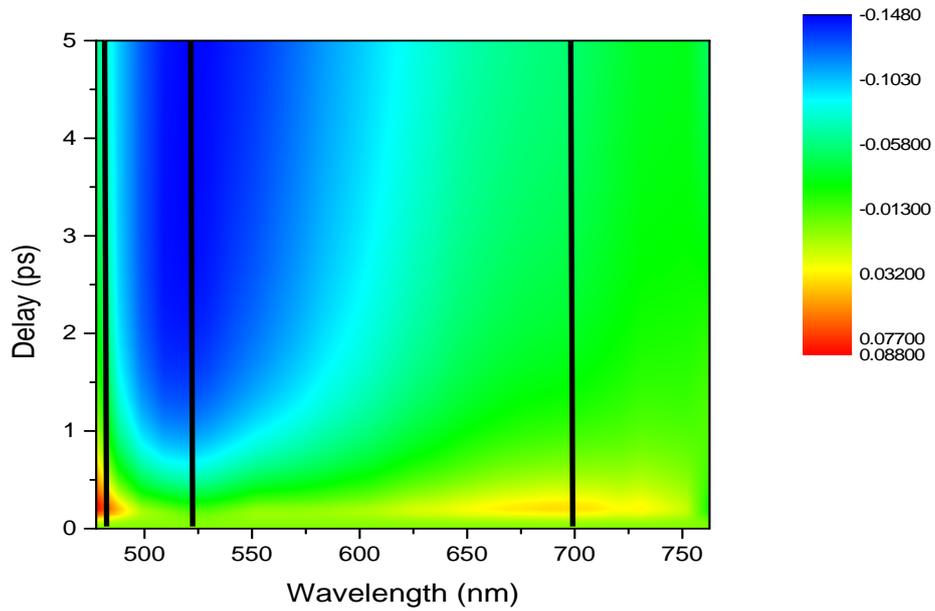

**Figure SI 16** Contour plot of ΔT/T for a PDI functionalized BP sample encapsulated in PMMA as an optically inert transparent matrix.

**SI 17:**

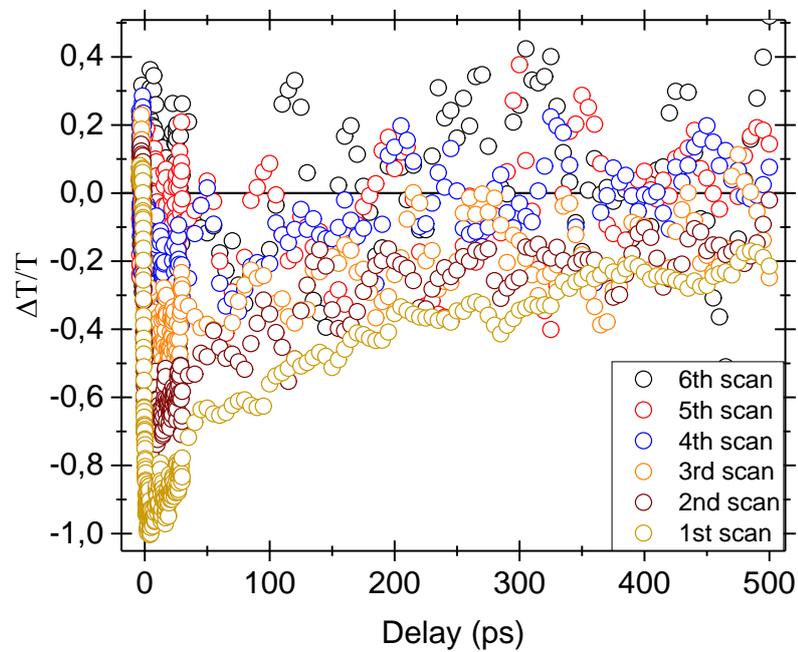

**Figure SI 17** Degradation of blank black phosphorous embedded in PMMA (dry sample) ($\lambda_{probe}$=562 nm)

**SI 18:**

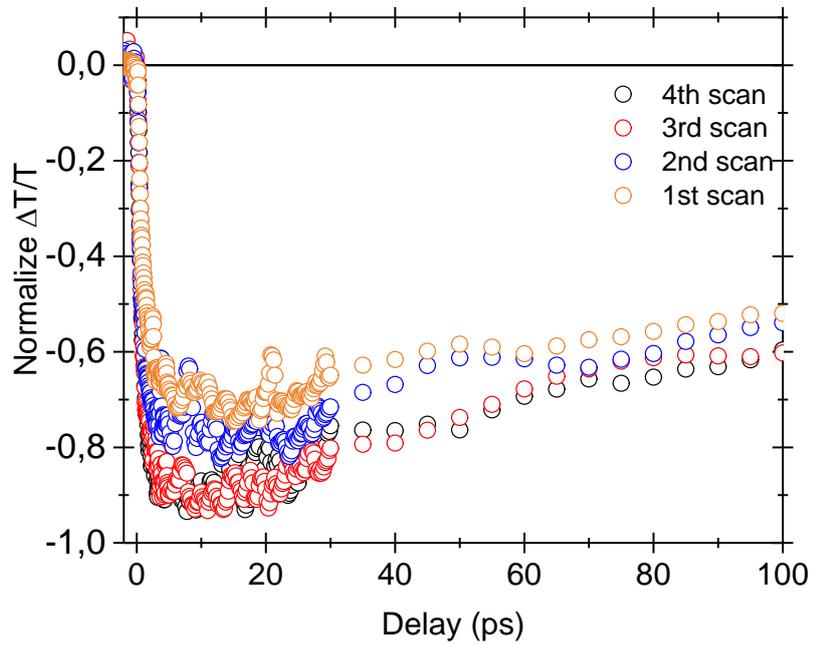

**Figure SI 18** Degradation of pdi black phosphorous embedded in PMMA (dry sample) ($\lambda_{probe}$=562 nm)

**SI 19:**

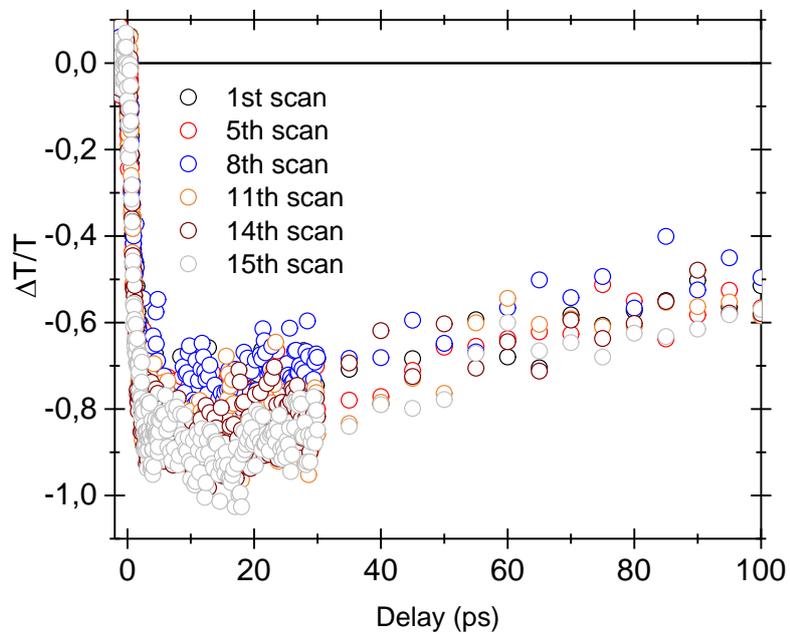

**Figure SI 19** Degradationn of pdi black phosphorous embedded in PMMA (liquid sample) ($\lambda_{probe}$=562 nm)

**SI 20:**

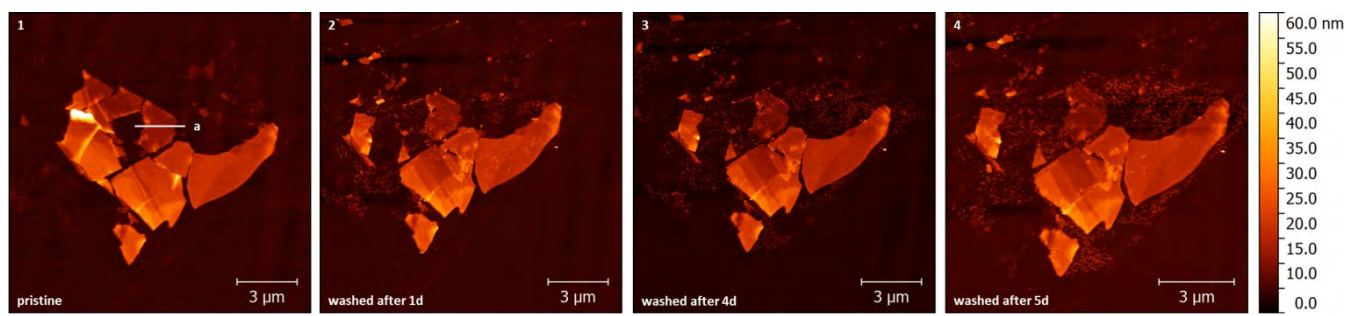

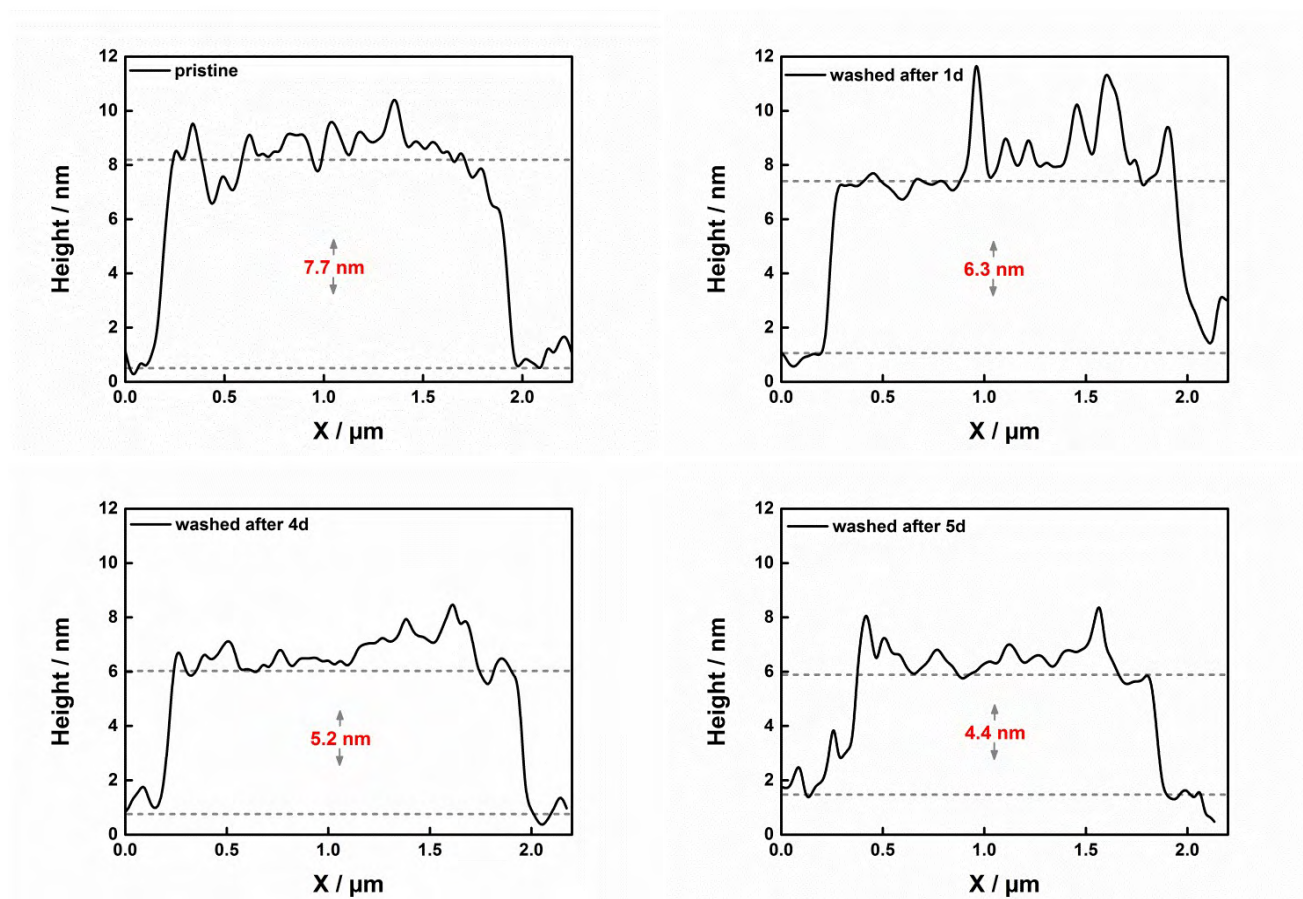

**Figure SI 20:**

(Top): AFM image of mechanically exfoliated FL-BP which has been washed 4 times with DI water to reduce the thickness. (Middle & Bottom): Corresponding AFM height profiles along line a illustrating the stepwise reduction of the thickness.

**SI 21:**

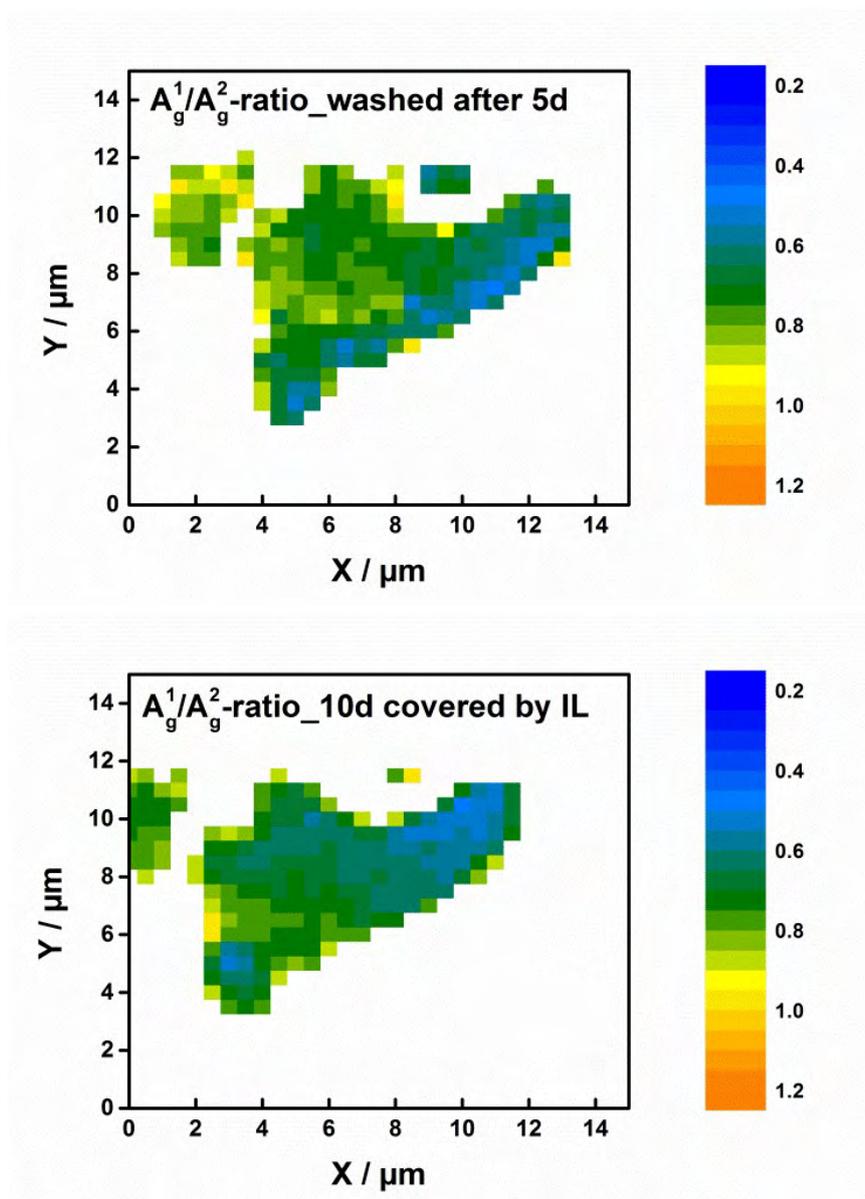

**Figure SI 21:**

$A^1_g/A^2_g$-ratio Raman mappings of the BP flakes shown in figure SI 8 after the last washing procedure with DI water (left) and after the removal of IL which covered the flakes for 10 days (right) in order to successfully prevent any degradation of the flakes.

**SI 22:**

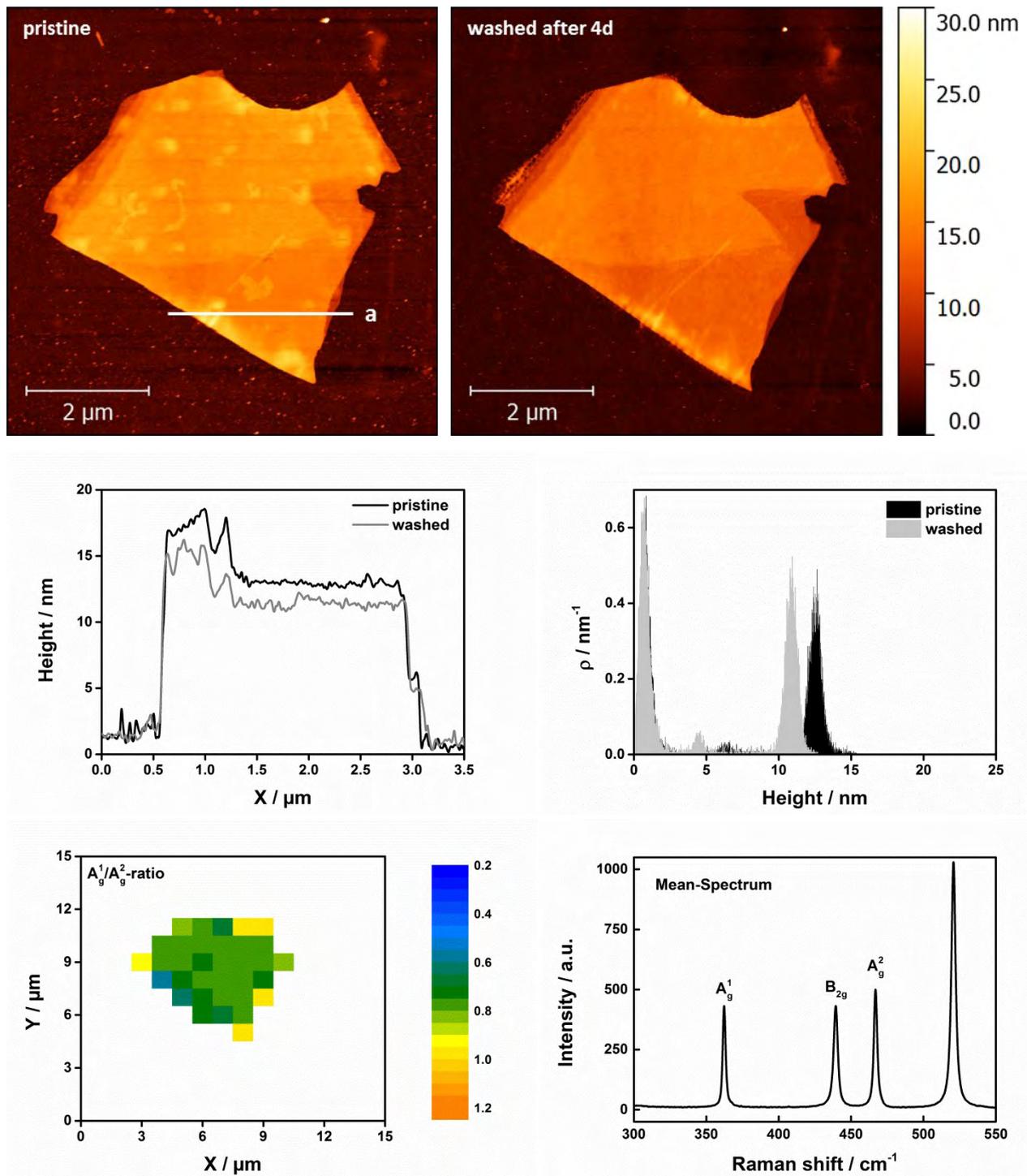

**Figure SI 22:**

(Top): AFM image of a mechanically exfoliated BP flake which has been washed with DI-water before it was electrically contacted to perform transport measurements. (Middle): Corresponding AFM height profiles along line a (left) and the statistical AFM evaluation (right) visualizing the reduced thickness of the flake. (Bottom): $A^1_g/A^2_g$-ratio Raman mapping of the BP flake after it was treated with DI-water and the related mean Raman spectrum showing the three characteristic vibrational modes of BP.

**SI 23:**

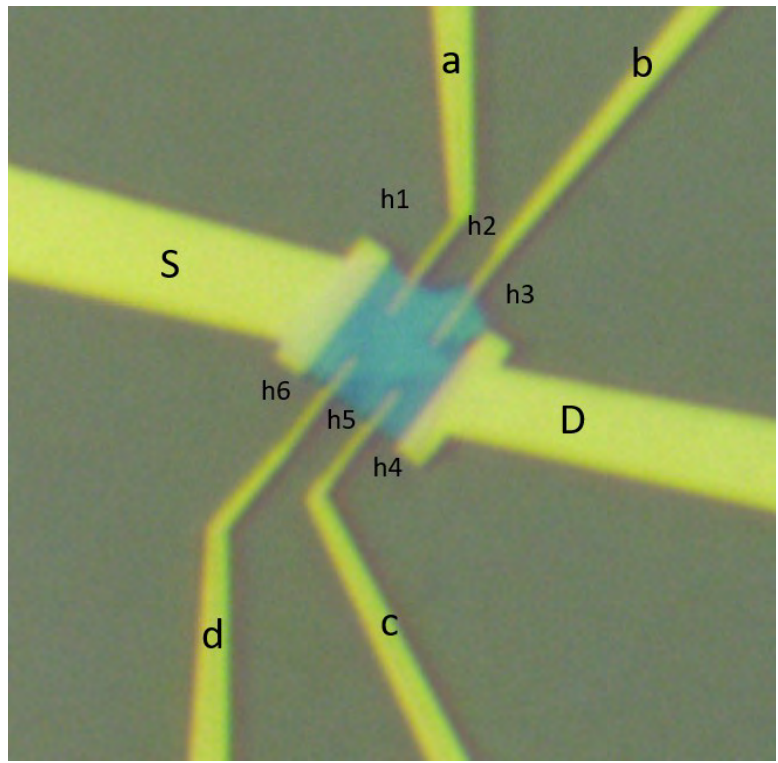

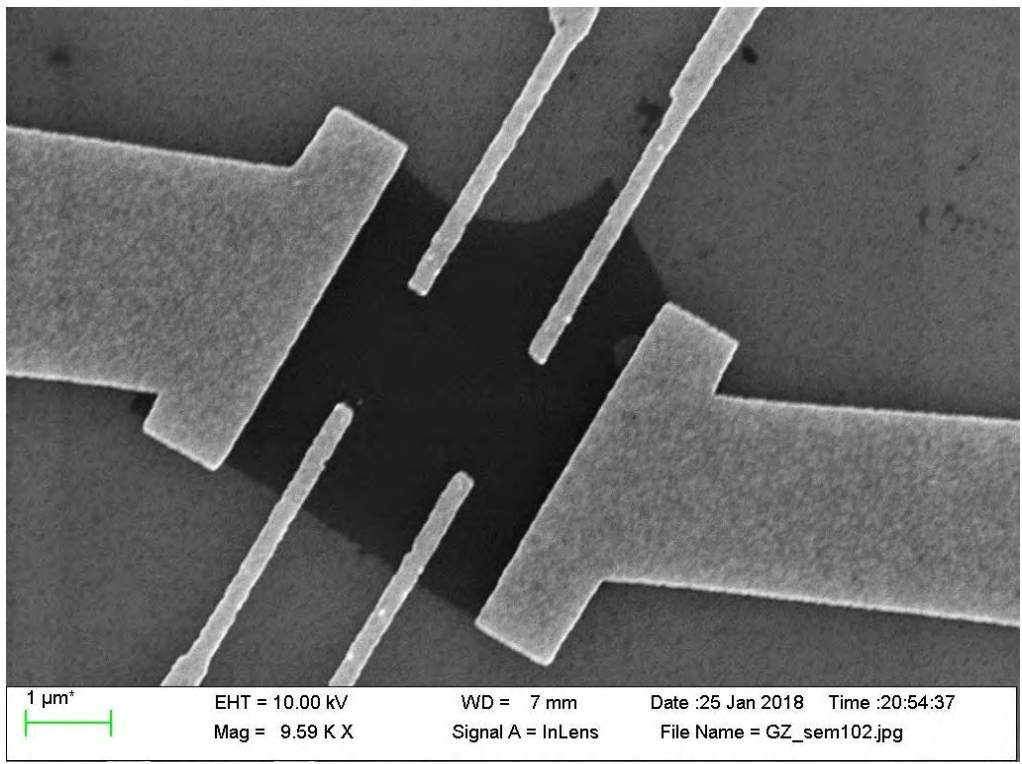

**Figure SI 23:**

(Top): Optical image of the rinsed BP flake which was contacted afterwards in a FET device. The mobility and carrier density were determined for the different heights of the flake. (Bottom): SEM-image of the rinsed BP flake.